\documentclass[a4paper, twoside, reqno, 12pt]{amsart}
\RequirePackage{fix-cm} 

\usepackage{fixltx2e}     

\usepackage{amssymb}
\usepackage{amsmath}
\usepackage{mathrsfs, dsfont}

\usepackage{a4wide}

\usepackage{acronym}

\usepackage{indentfirst}
\usepackage{graphicx}
\usepackage{psfrag}

\usepackage[usenames,dvipsnames,pdf]{pstricks}
\usepackage{epsfig}
\usepackage{pst-grad} 
\usepackage{pst-plot} 

\usepackage{subfigure}

\usepackage{longtable}

\usepackage[french,english]{babel}
\usepackage[latin1]{inputenc}

\usepackage{color}
\definecolor{oneblue}{rgb}{0,0.0,0.75}
\usepackage[colorlinks,
            urlcolor=oneblue,
            linkcolor=oneblue,
            citecolor=oneblue,
            bookmarksopen=false,
            pagebackref]{hyperref}

\numberwithin{equation}{section}

\newtheorem{remark}{Remark}

\newcommand{\uh}{\hat{u}}
\newcommand{\R}{\mathbb{R}}
\newcommand{\Z}{\mathbb{Z}}
\newcommand{\M}{\mathbb{M}}
\newcommand{\K}{\mathbb{K}}
\newcommand{\J}{\mathbb{J}}
\newcommand{\I}{\mathcal{I}}
\newcommand{\D}{\mathcal{D}}
\newcommand{\F}{\mathcal{F}}
\newcommand{\ud}{\mathrm{d}}
\newcommand{\Mm}{\mathcal{M}}
\renewcommand{\L}{\mathcal{L}}
\renewcommand{\S}{\mathcal{S}}
\renewcommand{\H}{\mathcal{H}}
\newcommand{\eps}{\varepsilon}

\newcommand{\z}{\boldsymbol{z}}
\newcommand{\N}{\ensuremath{\mathcal{N}}}
\renewcommand{\L}{\ensuremath{\mathcal{L}}}

\DeclareMathOperator{\sech}{sech}

\newcommand{\grad}{\boldsymbol{\nabla}}

\newcommand{\erfc}{\mathop{\mathrm{erfc}}}
\newcommand{\Span}{\mathop{\mathrm{span}}}
\renewcommand{\Pr}{\ensuremath{\mathbb{P}}}

\newcommand{\od}[2]{\frac{\mathrm{d}\,#1}{\mathrm{d}\/#2}}

\newcommand{\half}{{\textstyle{1\over2}}}
\newcommand{\sixth}{{\textstyle{1\over6}}}

\begin{document}

\title[Geometric schemes for the KdV equation]{Geometric numerical schemes for the KdV equation}

\author[D. Dutykh]{Denys Dutykh$^*$}
\address{LAMA, UMR 5127 CNRS, Universit\'e de Savoie, Campus Scientifique, 73376 Le Bourget-du-Lac Cedex, France}
\email{Denys.Dutykh@univ-savoie.fr}
\urladdr{http://www.lama.univ-savoie.fr/~dutykh/}
\thanks{$^*$ Corresponding author}

\author[M. Chhay]{Marx Chhay}
\address{LOCIE, FRE CNRS 3220, Universit\'e de Savoie, Campus Scientifique, 73376 Le Bourget-du-Lac Cedex, France}
\email{Marx.Chhay@univ-savoie.fr}

\author[F. Fedele]{Francesco Fedele}
\address{School of Civil and Environmental Engineering and School of Electrical and Computer Engineering, Georgia Institute of
Technology, Atlanta, USA}
\email{fedele@gatech.edu}
\urladdr{http://www.ce.gatech.edu/people/faculty/511/overview}

\begin{abstract}
Geometric discretizations that preserve certain Hamiltonian structures at the discrete level has been proven to enhance the accuracy of numerical schemes. In particular, numerous symplectic and multi-symplectic schemes have been proposed to solve numerically the celebrated Korteweg-de Vries (KdV) equation. In this work, we show that geometrical schemes are as much robust and accurate as Fourier-type pseudo-spectral methods for computing the long-time KdV dynamics, and thus more suitable to model complex nonlinear wave phenomena.
\end{abstract}


\maketitle

\tableofcontents

\section{Introduction}

The celebrated Korteweg - de Vries (KdV) equation has been extensively studied during the last 50 years. Its great scientific success is due to several reasons. First of all, this equation arises naturally in various fields ranging from hydrodynamics to plasma physics \cite{Lamb1980}. Moreover, it enjoys numerous nice mathematical properties \cite{Miura1976} such as a Hamiltonian and Lagrangian structures and integrability \cite{Zakharov1972a}, which yields exact analytical solitary and cnoidal wave solutions via the inverse scattering transform \cite{Osborne1995, Osborne2010}.

In this study we exploit the  KdV equation as a prototype model for dispersive wave (weak) turbulence in shallow waters. This subject has been extensively studied during the last 20 years \cite{Zakharov1992, Majda1997, Cai2001, Zakharov2001, Berger2003, Annenkov2006}, especially for deep water waves \cite{Zakharov1992}. For the KdV regime, the nonlinear evolution of a random sea of elementary waves (sea state) has been investigated in numerous studies using analytical and/or pseudo-spectral numerical methods (see, for example \cite{Osborne1993, Garnier2001, Salupere2003, Pelinovsky2006, Zahibo2009, Sergeeva2011}), but this list of references is far from being exhaustive. Moreover, some aspects of the stochastic KdV equation has been addressed in \cite{Debussche1999, Lin2006, DeBouard2007}. Note that  the KdV equation is integrable and the existence of an infinite number of integrals of motion prevents the excitation of temporal chaos (see, for example \cite{Cai2001}). Nevertheless we believe that studying the KdV dynamics is a good starting point to assess the real performance of (multi-)symplectic discretizations on realistic benchmarks stemming from weak turbulence theory \cite{Cai2001, Salupere2003, Pelinovsky2006}. There is also a closely connected subject related to solitonic gas. Dense solitonic gases can be studied using kinetic methods \cite{Zakharov1971, El2005a}. In this work, we will perform some simulations of a rarefied solitonic gas under the KdV (integrable) dynamics using direct numerical simulations \cite{Abdullaev1995}.

The numerical discretization of the KdV equation can be done by finite differences \cite{Taha1984, Schiesser1994, Ismail2000}, finite volumes \cite{Benkhaldoun2008, Dutykh2010e}, finite elements \cite{Arnold1980, Bona2007}, discontinuous Galerkin \cite{Levy2004} and, of course, by spectral methods \cite{Maday1988, Helal2001, Trefethen2000, Korkmaz2010}. However, recently the so-called geometrical numerical discretizations have been developed  \cite{McLachlan1993, Hairer2002, Lew2003, Leimkuhler2004, Kim2007a, Kim2008, Chhay2011}. The main idea behind these methods is to preserve some geometric properties of the continuous equation at the discrete level  \cite{Olver1993}. For instance, symplectic integrators  that preserve exactly the (discrete) symplectic form have been developed mostly for finite dimensional Hamiltonian systems \cite{McLachlan1993, Calvo1993, Sanz-Sera1997, Hairer2002, Leimkuhler2004, Ascher2004}. They possess excellent numerical properties since they are able to represent correctly the phase space of the simulated system over long-time integration and even if large time steps are used. 

More recently, the concept of multi-symplectic PDEs has been established and developed \cite{Marsden1998, Bridges2001}. The main idea is to treat the time and space variables on equal footing \cite{Moore2003a, Bridges2006}, while in Hamiltonian systems, for instance, the time variable is privileged with respect to the space. Based on this special structure, multi-symplectic schemes have been proposed for several PDEs including the KdV equation \cite{Zhao2000, Bridges2001, Moore2003, Wang2003, Ascher2004, Ascher2005, Islas2005, Bridges2006, Schober2008}. These schemes are specifically designed to preserve exactly a discrete multi-symplectic conservation law. Perhaps, one of the most popular schemes in this family is the Preissman box scheme. In particular, the 8-point scheme proposed by  \textsc{Ascher} \& \textsc{McLachlan} (2004) \cite{Ascher2004} will be used in our studies. Note that from a pragmatical point of view, it is not obvious what are the  advantages in using multi-symplectic schemes versus the symplectic ones. This question is partially addressed in this study. To do so, we will use certain symplectic and multi-symplectic schemes proposed earlier in \cite{Ascher2004, Ascher2005} in order to test their suitability to simulate complex interactions between radiative waves and coherent structures in a random wave field over a long evolution time. To date, geometric schemes have only been tested on simple academic benchmark problems  over short-to- intermediate evolution times. As a reference solution, we will use that obtained by a highly accurate pseudo-spectral method combined with a very high order time stepping. One of the aims of this study is to assess if the preservation of Hamiltonian properties allows the accurate modeling of more complex nonlinear phenomena than those of classical academic benchmarks. Indeed, the potential advantage of these geometric finite differences-based schemes is that they are able to simulated correctly the system dynamics even on coarse grids. Thus, simulations over potentially large domains of hundreds or even thousands of wavelengths may be feasibile beyond the standard spectral setting. Moreover, the solution is not restricted anymore to the class of periodic functions. Consequently, the effects of boundary conditions and confinement could be investigated. Further, the second aim is to investigate whether for Hamiltonian PDEs the canonical multi-symplectic discretization is more accurate than the classical symplectic scheme.

The present study is organized as follows. In Section \ref{sec:math} we introduce briefly the mathematical formulation of geometric schemes and their main properties. In next two subsections \ref{sec:sympl} and \ref{sec:msympl} we describe correspondingly the symplectic and multi-symplectic schemes in use. Subsection \ref{sec:spec} contains a brief description of a pseudo-spectral scheme, which will be used to compute the reference solution in the studied benchmark problems. Finally, the results of convergence tests and extensive comparisons are presented in Section \ref{sec:num}. The main conlcusions of this study are outlined in Section \ref{sec:concl}.

\section{Mathematical model and numerical schemes}\label{sec:math}

The one-dimensional KdV equation arises in many fields of science such as classical and magneto hydrodynamics, shallow water and stratified flows, internal waves, plasma physics and others \cite{Lamb1980}. In renormalized dimensionless variables, this equation can be written in the following simple form:
\begin{equation}\label{eq:kdv}
  u_t + uu_x + u_{xxx} = 0, \quad u(x,t): \R \times \R_+ \longrightarrow \R,
\end{equation}
where the subscripts $x$ and $t$ denote partial derivatives. In such setting, we interpret $u$ as the free surface elevation in shallow water hydrodynamics, but the simulations and results of this study are quite generic and can be interpreted in the context of other physical applications.

The integrability of  \eqref{eq:kdv}  \cite{Zakharov1972a} garanties the existence of an infinite number of invariants of motion. The first three invariants can be expressed as follows \cite{Dingemans1997}:
\begin{equation*}
  \I_1(t) = \int_\R u\,\ud x, \quad
  \I_2(t) = \int_\R u^2\,\ud x, \quad
  \I_3(t) = \int_\R \bigl[\sixth u^3 - \half (u_x)^2 \bigr]\,\ud x.
\end{equation*}
There exists a recurrence relation which allows to construct higher order invariants $\I_n$, for all $n>3$ \cite{Miura1968}:
\begin{equation*}
  \I_{n+1} = \I_{n+1}(\I_{n},\ldots,\I_{1}).
\end{equation*}

The invariant $\I_3$ has a special meaning since it is also a Hamiltonian for the KdV equation \cite{Olver1993}:
\begin{equation}\label{eq:ham}
  u_t = \J\frac{\delta\H}{\delta u}, \quad
  \J = -\partial_x, \quad
  \H = \I_3 = \int_\R \bigl[\sixth u^3 - \half (u_x)^2 \bigr]\,\ud x.
\end{equation}
This Hamiltonian structure sets the basis to construct a symplectic discretization for  \eqref{eq:kdv}.

Solitary and cnoidal type exact solutions are known to the KdV equation \cite{Dingemans1997}. In this article we will make extensive use of solitary waves in order to validate first the schemes, and then, to study the collective behaviour of the following localized solutions:
\begin{equation}\label{eq:soliton}
  u(x,t) = a\sech^2\bigl(\half\kappa(x-ct)\bigr), \quad a = 3c, \quad \kappa = \sqrt{c},
\end{equation}
where $c$ is the solitary wave propagation speed. Since the KdV equation is integrable, the solitary waves interact elastically and thus are solitons \cite{Zabusky1965}.

\subsection{Symplectic scheme}\label{sec:sympl}

In this and the next section we will follow in great lines the studies of \cite{Ascher2004, Bridges2006}. In order to obtain a symplectic scheme based on the Hamiltonian formulation \eqref{eq:ham}, we have to construct a semi-discretization in space. We choose the following discrete approximation of the Hamiltonian by the sum
\begin{equation*}
  \H_{\Delta x} = \Delta x\sum_{i}\biggl(\frac16 u_i^3 
  - \frac12\Bigl(\frac{u_{i+1} - u_i}{\Delta x}\Bigr)^2\biggr),
\end{equation*}
where $u_i := u(x_i,t)$, $x_i = i\Delta x, i\in\Z$ are the points of a uniform discretization in space for the sake of simplicity. Now we have a Hamiltonian system of coupled nonlinear ODEs, where the discretized version of the operator $\J_{\Delta x}$ is given by the central difference formula:
\begin{equation*}
  \od{u_i}{t} = \J_{\Delta x}\grad\H_{\Delta x}(u_i), \qquad
  \J_{\Delta x} = -\frac{\{\cdot\}_{i+1} - \{\cdot\}_{i-1}}{2\Delta x}.
\end{equation*}
These equations can be expanded in component-wise form needed for the practical implementation:
\begin{eqnarray*}
  \od{u_i}{t} &=& -\frac{\half u_{i+1}^2 - \half u_{i-1}^2}{2\Delta x} 
  -\frac{1}{2\Delta x^2}\Bigl(\frac{u_{i} - 2u_{i+1} + u_{i+2}}{\Delta x^2} - \frac{u_{i-2} - 2u_{i-1} + u_{i}}{\Delta x^2}\Bigr) \\
  &=& -\frac{u_{i+1}^2 - u_{i-1}^2}{4\Delta x} -\frac{1}{2\Delta x^3}\bigl(-u_{i-2} + 2u_{i-1} - 2u_{i+1} + u_{i+2}\bigr).
\end{eqnarray*}
Finally, in order to obtain a fully discrete scheme we use the midpoint rule in time, which provides us a symplectic integrator:
\begin{equation}\label{eq:midpoint}
  \frac{u_i^{(n+1)} - u_i^{(n)}}{\Delta t} = \J_{\Delta x}\grad\H_{\Delta x}\Bigl(\frac{u_i^{(n+1)} + u_i^{(n)}}{2}\Bigr).
\end{equation}
The last system of equations is implicit and thus, it has to be solved at each time step. To do so, we employ the classical Newton iteration \cite{Isaacson1966} which converges very quickly in practice since the nonlinearity is only quadratic and a good initial approximation to the solution can be obtained with a simple inexpensive explicit scheme.

\subsection{Multi-symplectic scheme}\label{sec:msympl}

Additionally to the Hamiltonian formulation, the KdV equation \eqref{eq:kdv} possesses also a multi-symplectic canonical structure:
\begin{equation}\label{eq:ms}
  \M\z_t + \K\z_x = \grad_{\z}\S(\z),
\end{equation}
where $\z = {}^t(\phi, u, v, w)$ is the vector of dependent variables and skew-symmetric matrices $\M$ and $\K$ along with the multi-symplectic Hamiltonian functional $\S(\z)$ are defined as
\begin{equation*}
  \M = \begin{pmatrix}
     0       &  \frac12 & 0 & 0 \\
    -\frac12 &    0     & 0 & 0 \\
     0       &    0     & 0 & 0 \\
     0       &    0     & 0 & 0
  \end{pmatrix}, \quad
  \K = \begin{pmatrix}
    0 &  0 & 0 & 1 \\
    0 &  0 &-1 & 0 \\
    0 &  1 & 0 & 0 \\
   -1 &  0 & 0 & 0
  \end{pmatrix}, \quad
  \S(\z) = \frac12 v^2 - uw + \frac16 u^3.
\end{equation*}
Rewriting in component-wise way, the multi-symplectic formulation \eqref{eq:ms} yields the following set of equations
\begin{equation*}
  u = \phi_x, \qquad v = u_x, \qquad w = \frac12\phi_t + v_x + \frac12 u^2, \qquad \frac12 u_t + w_x = 0.
\end{equation*}
\noindent
The main geometrical property of the Hamiltonian PDEs formulation \eqref{eq:ms} is given by the multi-symplectic local conservation law:
\begin{equation*}
  \omega_t + \kappa_x = 0, \quad \omega := \frac12\, \ud\z \wedge \M\ud\z,
  \quad \kappa := \frac12\, \ud\z \wedge \K\ud\z,
\end{equation*}
where $\omega$ and $\kappa$ are pre-symplectic forms associated to the time and space directions correspondingly:
\begin{equation*}
  \omega = \half \ud\phi \wedge \ud u, \qquad
  \kappa = \ud\phi \wedge \ud w + \ud v\wedge \ud u.
\end{equation*}
\noindent
If the multi-symplectic Hamiltonian function $\S(\z)$ does not depend explicitly on time and space variables $x$ and $t$ (as it is the case here), energy and momentum are locally conserved:
\begin{equation*}
 \begin{array}{rrllrll}
  E_t + F_x = 0, & \quad E(\z) &:=& \S(\z) - \frac12\langle\K\z_x,\z\rangle, & \quad F(\z) &:=& \frac12\langle\K\z_t,\z\rangle, \\
  I_t + G_x = 0, & \quad I(\z) &:=& \frac12\langle\M\z_x,\z\rangle, & \quad G(\z) &:=& \S(\z) - \frac12\langle\M\z_t,\z\rangle.
 \end{array}
\end{equation*}
In our particular case, the generalized energy, momentum and their fluxes take corresponding the following form:
\begin{eqnarray*}
  E(\z) &=& \frac12 v^2 - uw + \frac16 u^3 - \frac12\bigl(w_x\phi - \phi_x w - v_xu + vu_x\bigr), \\
  F(\z) &=& \frac12\bigl(w_t\phi - \phi_t w - v_t u + vu_t\bigr), \\
  I(\z) &=& \frac14\bigl(-\phi_x u + \phi u_x\bigr), \\
  G(\z) &=& \frac12 v^2 - uw + \frac16 u^3 - \frac14\bigl(-\phi_tu + \phi u_t\bigr).
\end{eqnarray*}
\noindent

\begin{remark}
The multi-symplectic formulation \eqref{eq:ms} can be formally obtained from the following Lagrangian functional $L$ by applying the Hamilton's principle \cite{Hydon2005}:
\begin{equation*}
  L = \int\,\ud t\int \L\,\ud x, \qquad
  \L = \frac{1}{2}\langle\M\z_t,\z\rangle + 
  \frac{1}{2}\langle\K\z_x,\z\rangle - \S(\z).
\end{equation*}
\end{remark}
\noindent
Using the presented above multi-symplectic structure \eqref{eq:ms} for the KdV equation \eqref{eq:kdv} one can construct various multi-symplectic discretizations \cite{Bridges2006}. Recall that a numerical algorithm is said to be multi-symplectic when it preserves exactly a discrete formulation of the above multi-symplectic conservation law. In this study we choose a multi-symplectic scheme as derived in \cite{Ascher2004}:
\begin{equation}\label{eq:p8box}
  \D_t\Mm_x^3 u_i^{(n)} + \frac12\D_x\Mm_x\bigl(\Mm_t\Mm_x u_i^{(n)}\bigr)^2 + \Mm_t\D_x^3 u_i^{(n)} = 0, \qquad i\in\Z, \quad n\in\Z_+,
\end{equation}
where the discrete difference and averaging operators are defined as
\begin{eqnarray*}
  \D_x u_i^{(n)} &=& \frac{u_{i+1}^{(n)} - u_{i}^{(n)}}{\Delta x}, \quad
  \D_t u_i^{(n)} = \frac{u_{i}^{(n+1)} - u_{i}^{(n)}}{\Delta t}, \\
  \Mm_x u_i^{(n)} &=& \frac{u_{i}^{(n)} + u_{i+1}^{(n)}}{2}, \quad
  \Mm_t u_i^{(n)} = \frac{u_{i}^{(n)} + u_{i}^{(n+1)}}{2}.
\end{eqnarray*}
The finite difference scheme \eqref{eq:p8box} is called the 8-point box scheme since it involves precisely 8 points in its stencil distributed on two temporal layers. This scheme was already shown to have promising numerical properties in the original study by \textsc{Ascher} \& \textsc{McLachlan} (2004) \cite{Ascher2004}. Partially, its success can be explained from the linear dispersion analysis. Namely, it was shown in \cite{Ascher2004} that the box schemes preserve qualitatively the dispersion relation of any linear multy-symplectic PDE. The behaviour of this scheme in highly-nonlinear situations will be investigated below.

\subsection{Spectral method}\label{sec:spec}

In order to assess the accuracy of the above geometric discretizations in several tests, we need a reference exact solution for each benchmark problem. This will be provided by a highly accurate Fourier-type pseudo-spectral method that we will briefly describe below \cite{Trefethen2000, Boyd2000, Canuto2006}. 

Denote by $\uh(k,t) = \F\{u\}$ the Fourier transform of $u(x,t)$ in $x$, where $k$ is the wavenumber. Then, by Fourier-transforming the KdV equation \eqref{eq:kdv} yields
\begin{equation}\label{eq:sp}
  \uh_t - ik^3\uh = -\half ik\widehat{(u^2)}.
\end{equation}
Here, the spatial derivatives are computed in spectral space, whereas the nonlinear product is computed in real space and dealised using the classical $3/2$th rule. The overall implementation is very efficient thanks to the FFT algorithm \cite{FFTW98, Frigo2005}. In order to improve the time-stepping we will use the so-called integrating factor technique. This consists of the exact integration of the linear terms of \eqref{eq:sp}, viz. 
\begin{equation}\label{eq:spectral}
  \hat{v}_t = e^{(t-t_0)\L}\cdot\N\Bigl\{e^{-(t-t_0)\L}\cdot\hat{v}\Bigr\},
   \qquad 
  \hat{v}(t) \equiv e^{(t-t_0)\L}\cdot\uh(t), \qquad 
  \hat{v}(t_0) = \uh(t_0),
\end{equation}
where the linear and nonlinear operators $\L$ and $\N$ are defined through their symbols as
\begin{equation*}
  \L := -ik^3, \qquad \N := -\half ik\widehat{(u^2)}.
\end{equation*}
This allows to increase substantially the accuracy and the stability region of the time marching scheme (see, for example \cite{Trefethen2000}). Finally, the resulting system of ODEs is discretized in time by the Verner's embedded adaptive 9(8) Runge--Kutta scheme \cite{Verner1978}. The time step is adapted automatically according to the H211b digital filter approach \cite{Soderlind2003, Soderlind2006}.

\section{Numerical results}\label{sec:num}

To assess if symplectic and multi-symplectic discretizations can be exploited to simulate efficiently complex nonlinear wave phenomena, we will test the schemes \eqref{eq:midpoint}, \eqref{eq:p8box} and \eqref{eq:spectral} in several benchmark problems. The numerical solution of the pseudo-spectral scheme will serve as the exact reference solution.

\subsection{Convergence tests}

Our first numerical test is designed to validate the implementation of the two geometric schemes and their order of convergence to the exact solution.

The classical CFL condition proposed for the first time by R.~\textsc{Courant} \emph{et al.} \cite{Courant1928, Courant1967} is defined as:
\begin{equation*}
  \mathrm{CFL} := \frac{\Delta t}{\Delta x}, \qquad
  \Delta t = \mathrm{CFL}\cdot\Delta x.
\end{equation*}
Note that in our study this condition is not dictated by any stability constraints. It is simply the ratio between time and space steps. Actually, the adopted geometric schemes are fully implicit and so unconditionally stable. Even more, our numerical tests show that reliable results can be obtained with CFL numbers equal to $8\ldots 9$. Perhaps, the best trade off between quality (accuracy) of results and CPU time is attained for CFL numbers in the range $2 \ldots 2.5$. In our numerical experiments we used CFL $= 0.1$ to achieve accuracy.

\begin{table}
  \centering
  \begin{tabular}{c|c}
  \hline\hline
  \textit{Parameter} & \textit{Value} \\
  \hline
  Soliton speed, $c$ & 0.1 \\
  Soliton amplitude, $a = 3c$ & 0.3 \\
  CFL number & 0.1 \\
  Tolerance parameter & $10^{-13}$ \\
  Number of points, $N$ & $2^6\ldots 2^{12}$ \\
  Half-length of the domain, $[-\ell, \ell]$ & 80.0 \\
  Spatial discretization step, $\Delta x$ & $2.5\ldots 0.0390625$ \\
  Final simulation time, $T$ & $15.0$ \\
  \hline\hline
  \end{tabular}
  \vspace{0.5em}
  \caption{Geometrical schemes: numerical parameters used to study their convergence properties.}
  \label{tab:conv}
\end{table}

\begin{figure}
  \centering
  \subfigure[$L_2$ error]%
  {\includegraphics[width=0.49\textwidth]{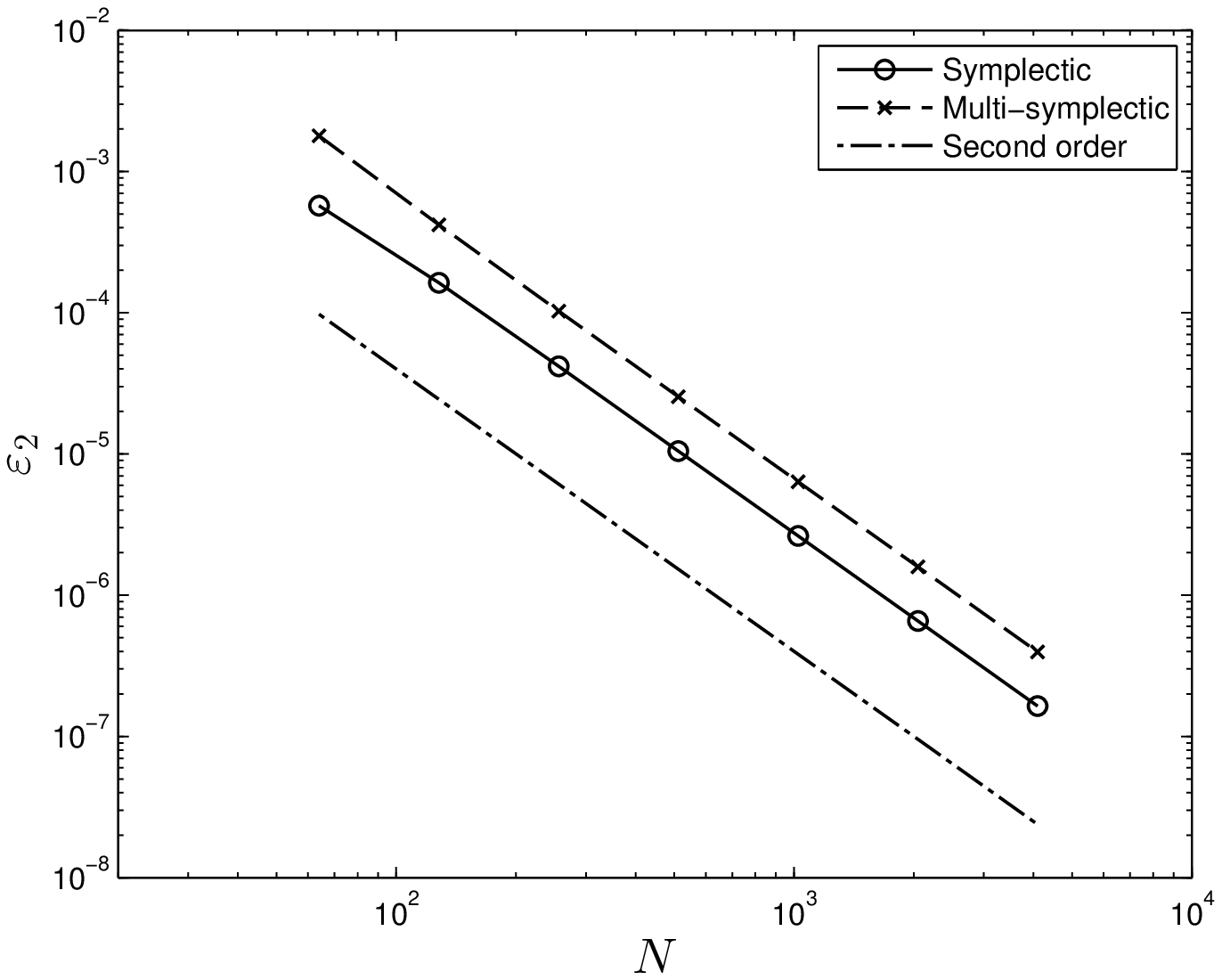}}
  \subfigure[$L_\infty$ error]%
  {\includegraphics[width=0.49\textwidth]{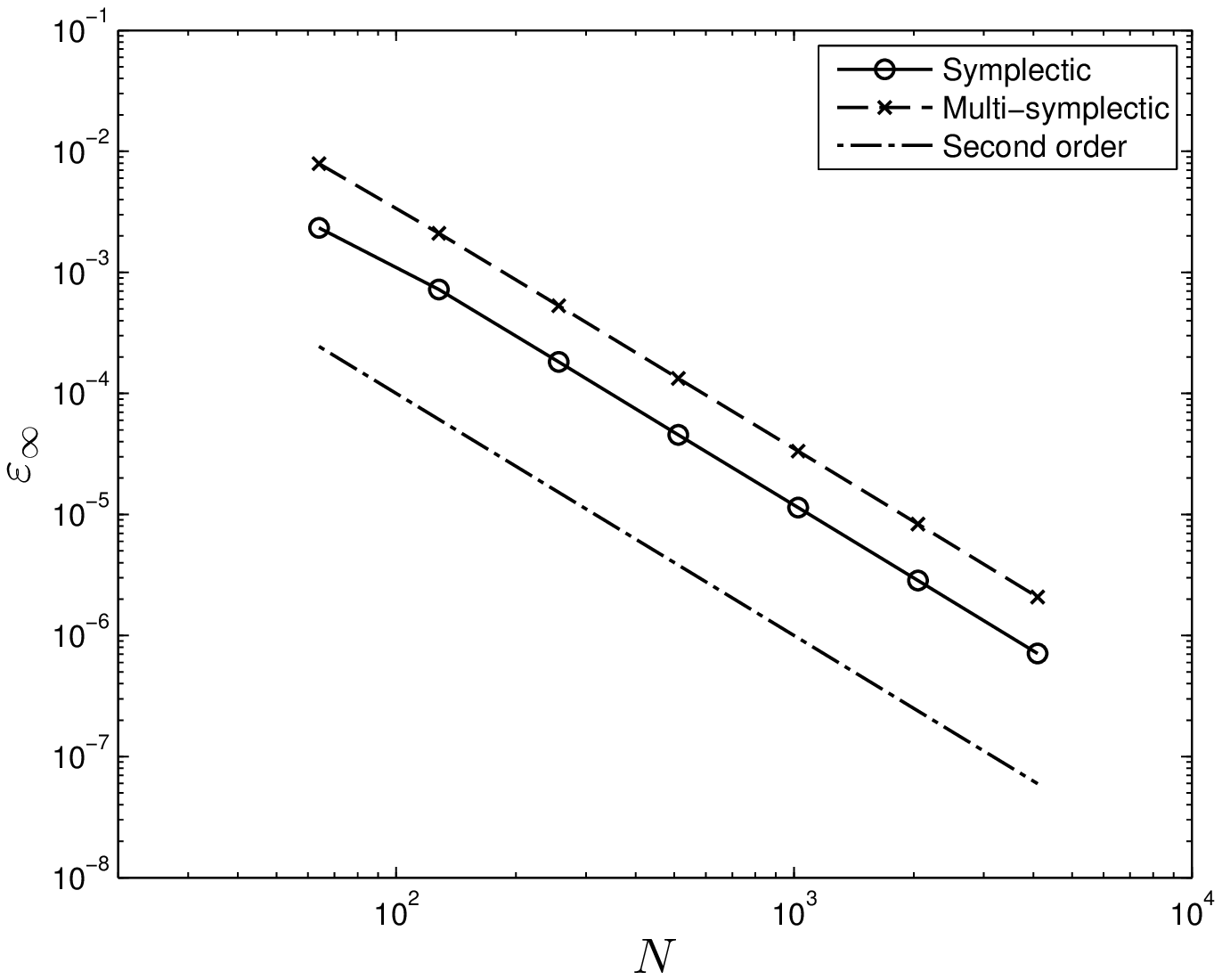}}
  \caption{Test-case A (single solitary wave): (Left) convergence error measured in $L_2$-norm and (right) in $L_\infty$-norm.}
  \label{fig:Err2Inf}
\end{figure}

\begin{figure}
  \centering
  \subfigure[ ]%
  {\includegraphics[width=0.49\textwidth]{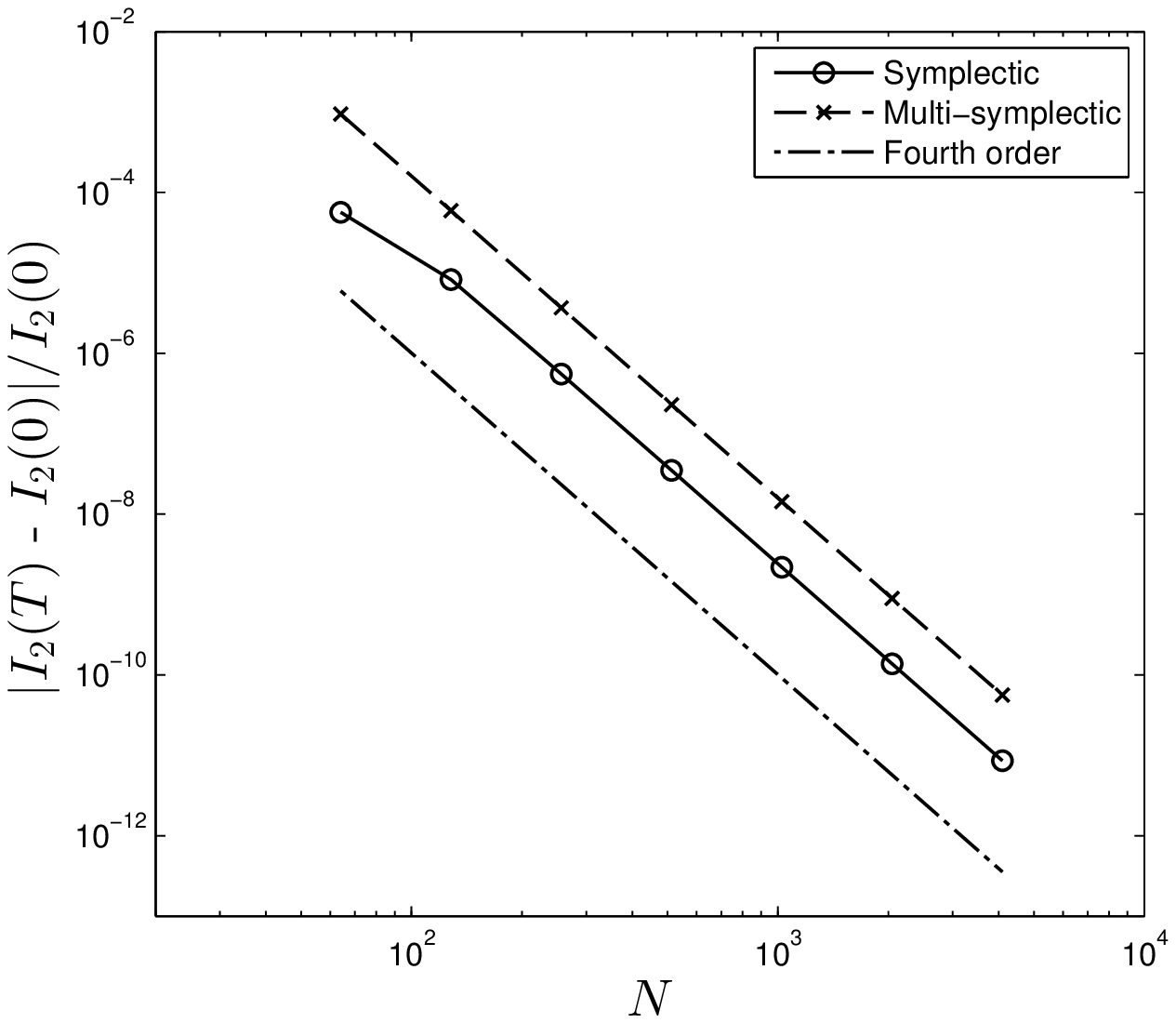}}
  \subfigure[ ]%
  {\includegraphics[width=0.49\textwidth]{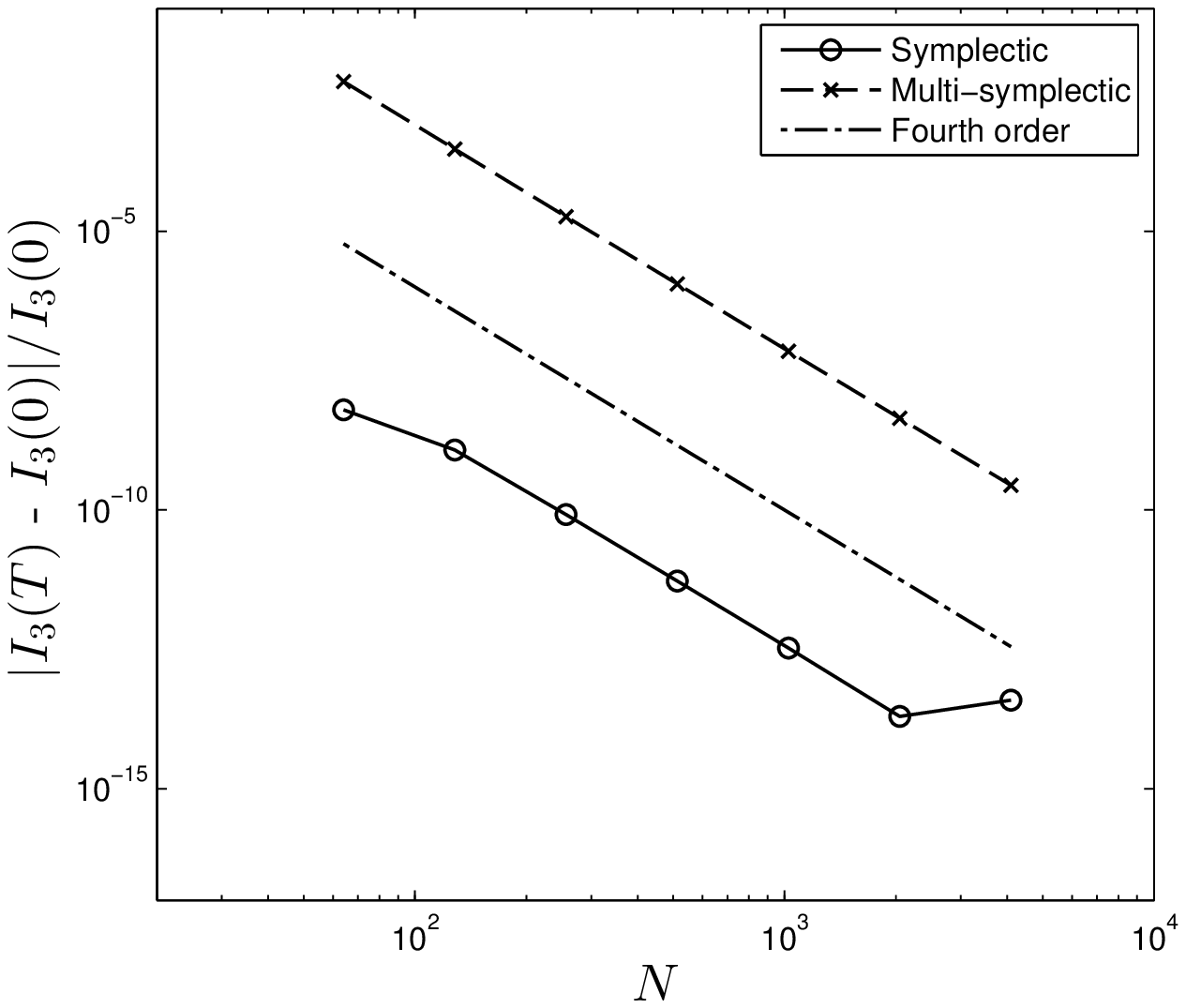}}
  \caption{Test-case A (single solitary wave): conservation of the invariants  $\I_2$ and $\I_3$ as function of the grid size $N$: the fourth order convergence rate.}
  \label{fig:ErrInv}
\end{figure}

\begin{figure}
  \centering
  \subfigure[ ]%
  {\includegraphics[width=0.49\textwidth]{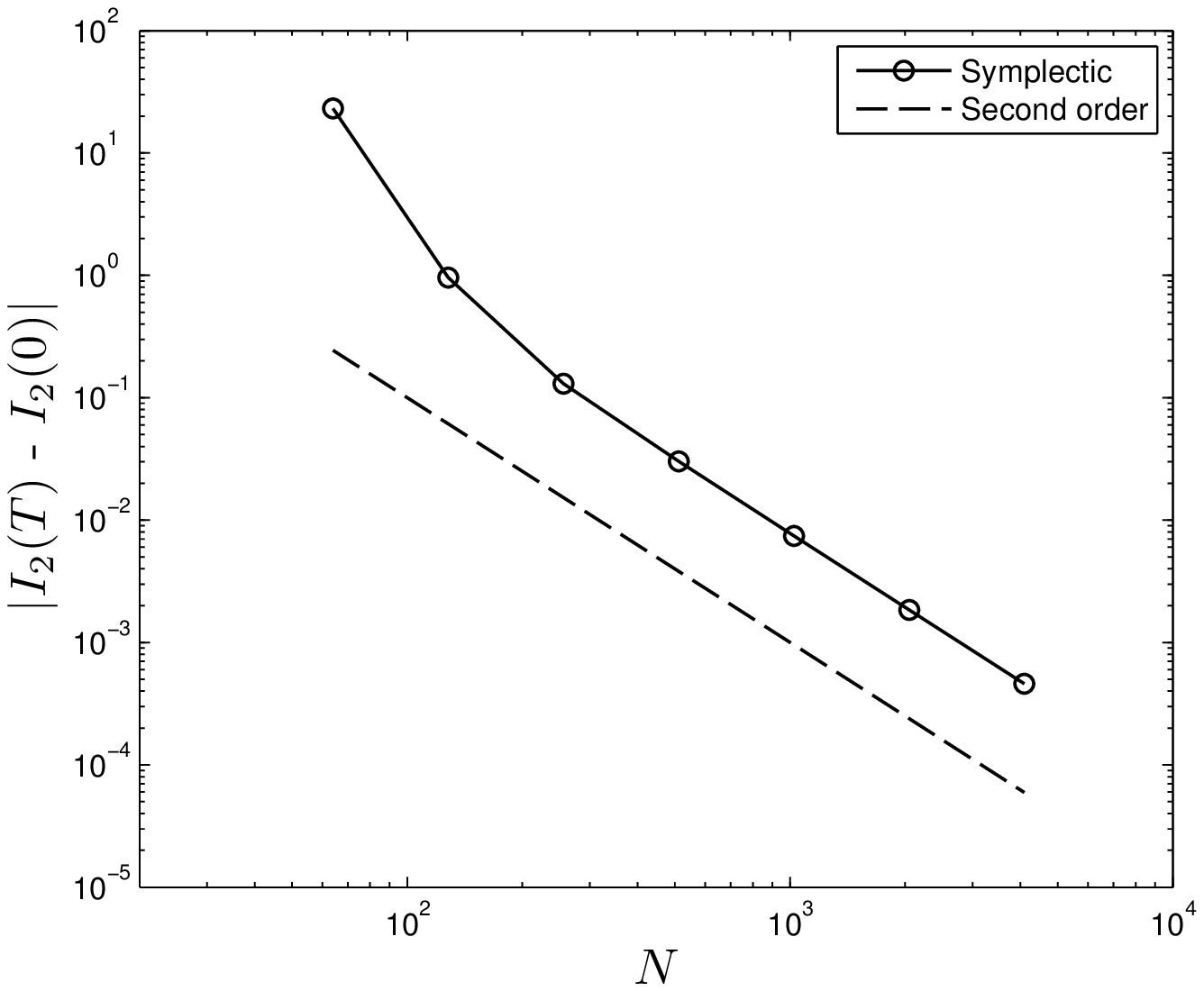}}
  \subfigure[ ]%
  {\includegraphics[width=0.49\textwidth]{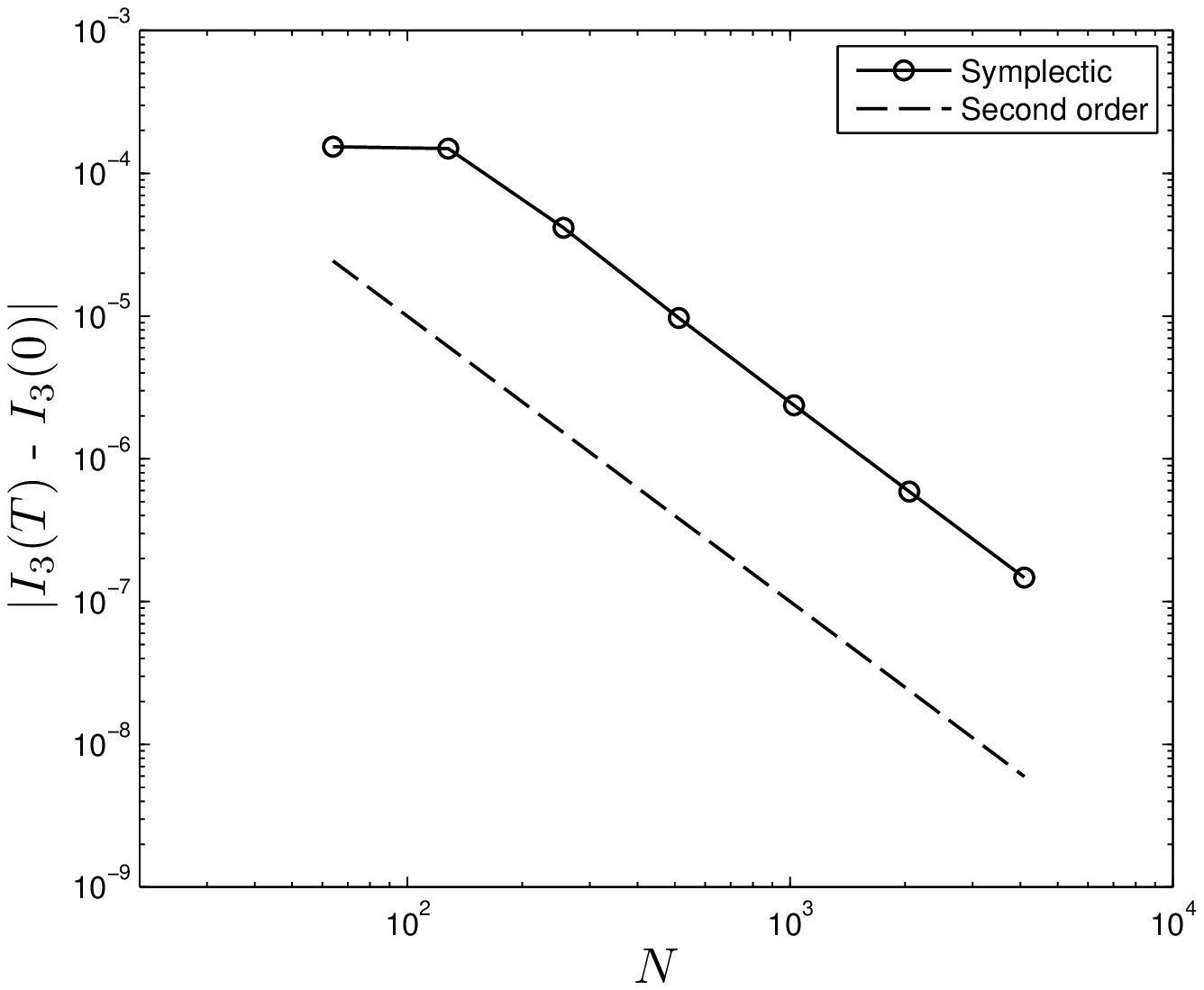}}
  \caption{Test-case B (two-soliton collision): conservation of the invariants $\I_2$ and $\I_3$ as function of the grid size $N$ for the symplectic scheme \eqref{eq:midpoint}: the second order convergence rate.}
  \label{fig:overtake}
\end{figure}

\begin{figure}
  \centering
  \includegraphics[width=0.69\textwidth]{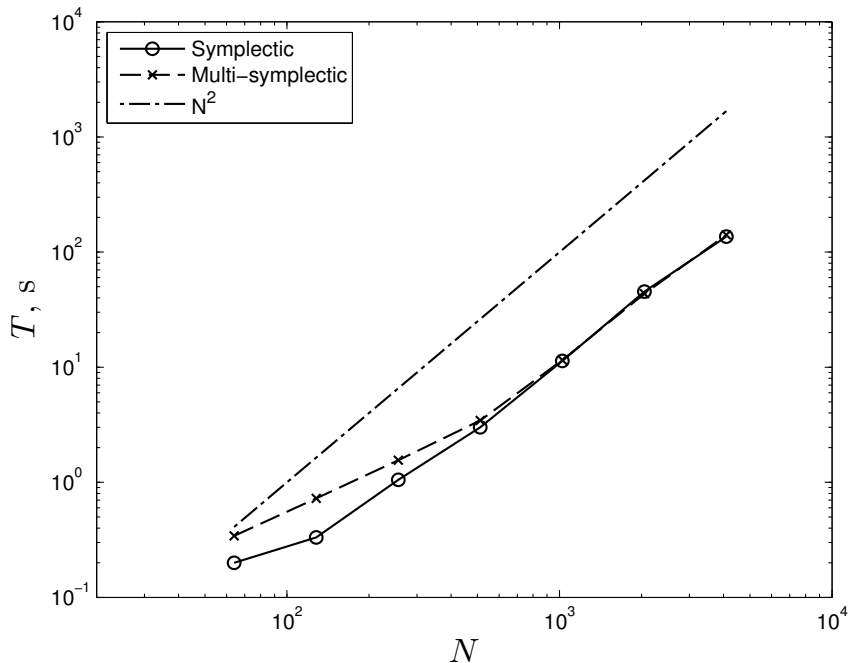}
  \caption{Test-case A (single solitary wave): observed CPU time as a function of the grid size $N$.}
  \label{fig:CPU}
\end{figure}

The first test-case (A) consists of a single soliton \eqref{eq:soliton} that freely propagates during some time $T$. Table \ref{tab:conv} reports the parameters adopted in the simulations. For a given grid of size $N$, at the final time $T$ we estimate the error between the numerical and exact solutions in both discrete $L_2$ and $L_\infty$ norms as function of $N$.   The computational results are presented in Figure \ref{fig:Err2Inf}. One can clearly observe that the theoretical second order convergence rate is attained in the numerical simulations. More precisely, the estimated rates are reported in Table \ref{tab:rates}. Invariants $\I_2$ and $\I_3$ show even better numerical properties as illustrated in Figure \ref{fig:ErrInv}. Namely, a super-convergent rate $\sim 4.0$ is observed in the conservation of non-trivial invariants. In particular, the symplectic scheme \eqref{eq:midpoint} even attains machine accuracy level for $\I_3$. This surprising excellent performance is due to the special solution that is simulated, viz. a travelling wave that propagates in space without changing shape under the KdV dynamics. For a general initial condition the theoretical second order rate will be recovered. Indeed, for the symplectic scheme \eqref{eq:midpoint} Figure \ref{fig:overtake} shows the error observed in the conservation of invariants for a second test-case (B) of the collision of two solitary waves (speeds $c_1 = 0.6$ and $c_2 = 0.1$, respectively). Finally, in Figure \ref{fig:CPU} we show the measured CPU time needed to carry out the computations as function of the number of discretization points $N$, i.e. grid size. Clearly, both symplectic and multi-symplectic schemes have similar asymptotic behaviour and the algorithmic complexity scales as $O(N^2)$ as expected. Indeed, at each time step we solve a finite number of sparse linear systems resulting from the Newton method applied to the nonlinear implicit discrete equations. In average this step can be done efficiently in roughly $O(N)$ number of operations, and we make $O(N)$ time steps $\Delta t$ to compute the solution at the final time $T$. Consequently, the optimal complexity is $O(N^2)$ as observed in our simulations.

\begin{remark}
Note that due to the conservative form of the geometric schemes, the invariant $\I_1$ is \emph{always} conserved at the machine precision level, and it does not depend on the number of grid points, time step, etc. Consequently, hereafter we omit to report this quantity in our results.
\end{remark}

\begin{table}
  \centering
  \begin{tabular}{|c||c|c|}
    \hline
    Norm             &   $L_2$ & $L_\infty$ \\
    \hline\hline
    Symplectic       & $1.972$ & $1.964$ \\
    Multi-symplectic & $2.019$ & $1.987$ \\
    \hline
  \end{tabular}
  \vspace{1em}
  \caption{Geometric schemes: estimated rates of convergence (results from Figure \ref{fig:Err2Inf}).}
  \label{tab:rates}
\end{table}

\begin{figure}
  \centering
  \includegraphics[width=0.69\textwidth]{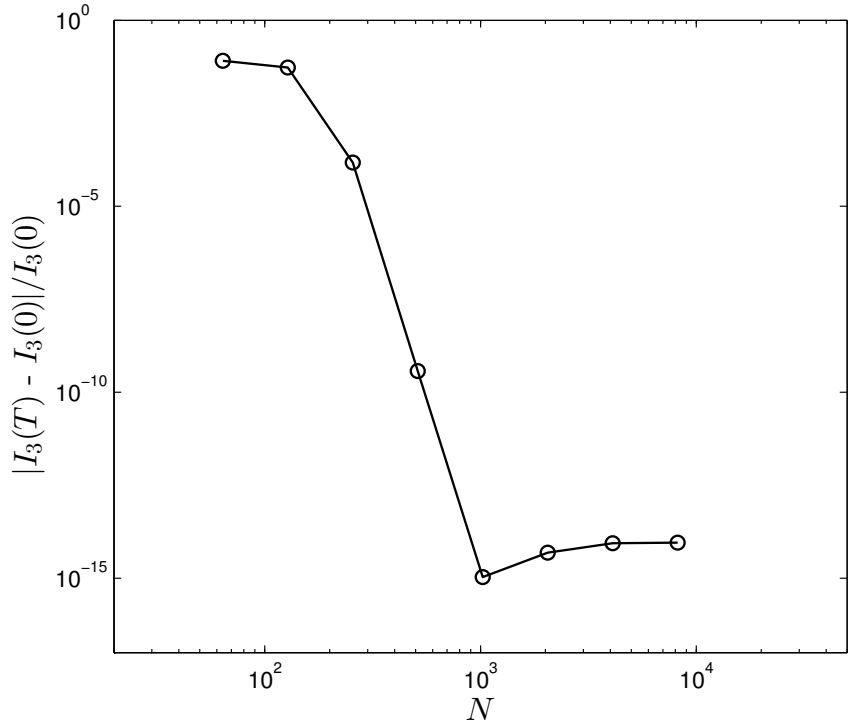}
  \caption{Test-case A (single solitary wave): convergence error for the invariant $\I_3$ using the pseudo-spectral scheme \eqref{eq:spectral}.}
  \label{fig:specI3}
\end{figure}

The pseudo-spectral discretization attains exponential convergence for sufficiently smooth solutions. For test-case A, in Figure \ref{fig:specI3} we report the observed error in the conservation of the invariant $\I_3$ as function of the number of Fourier modes $N$. Clearly, the error decays at an exponential rate down to the machine accuracy level ($\eps \sim 10^{-15}$) at roughly $N = 1000$, and it remains roughly constant as the grid is further refined. Hereafter, in the pseudo-spectral simulations we use $N = 4096$ to solve the KdV equation \eqref{eq:kdv}.

\subsection{Solitonic gas dynamics}

Consider as an initial sea state that of a random ensemble of solitons, viz. a solitonic gas. Assume that the soliton speeds are stochastically independent and Gaussian, that is
\begin{equation*}
  c_j \sim \N(c_0, \sigma^2), \quad j = 1,\ldots, M,
\end{equation*}
and the associated amplitudes $a_j$, $j = 1,\ldots, M$ follow from \eqref{eq:soliton}, $M$ being the number of solitons. The solitons are set initially as equally spaced and well separared ($\delta x \gg$ soliton width):
\begin{equation*}
  u(x,0) = \sum_{j=1}^M a_j\sech\bigl(\half\sqrt{c_j}(x - x_0 - j\delta x)\bigr)^2, \qquad a_j = 3c_j.
\end{equation*}

\begin{figure}
  \centering
  \subfigure[Complete view]%
  {\includegraphics[width=0.49\textwidth]{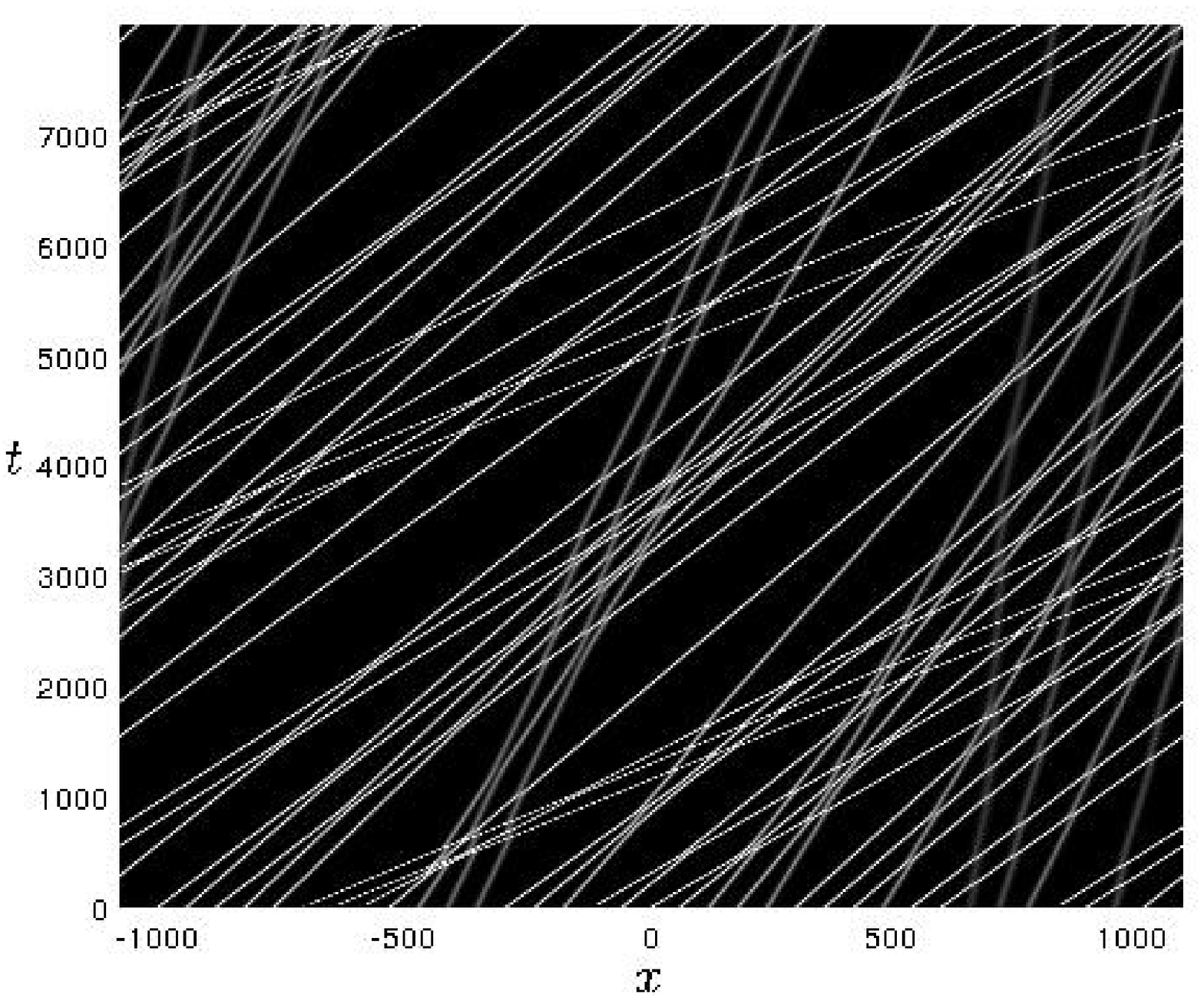}}
  \subfigure[Zoom]%
  {\includegraphics[width=0.49\textwidth]{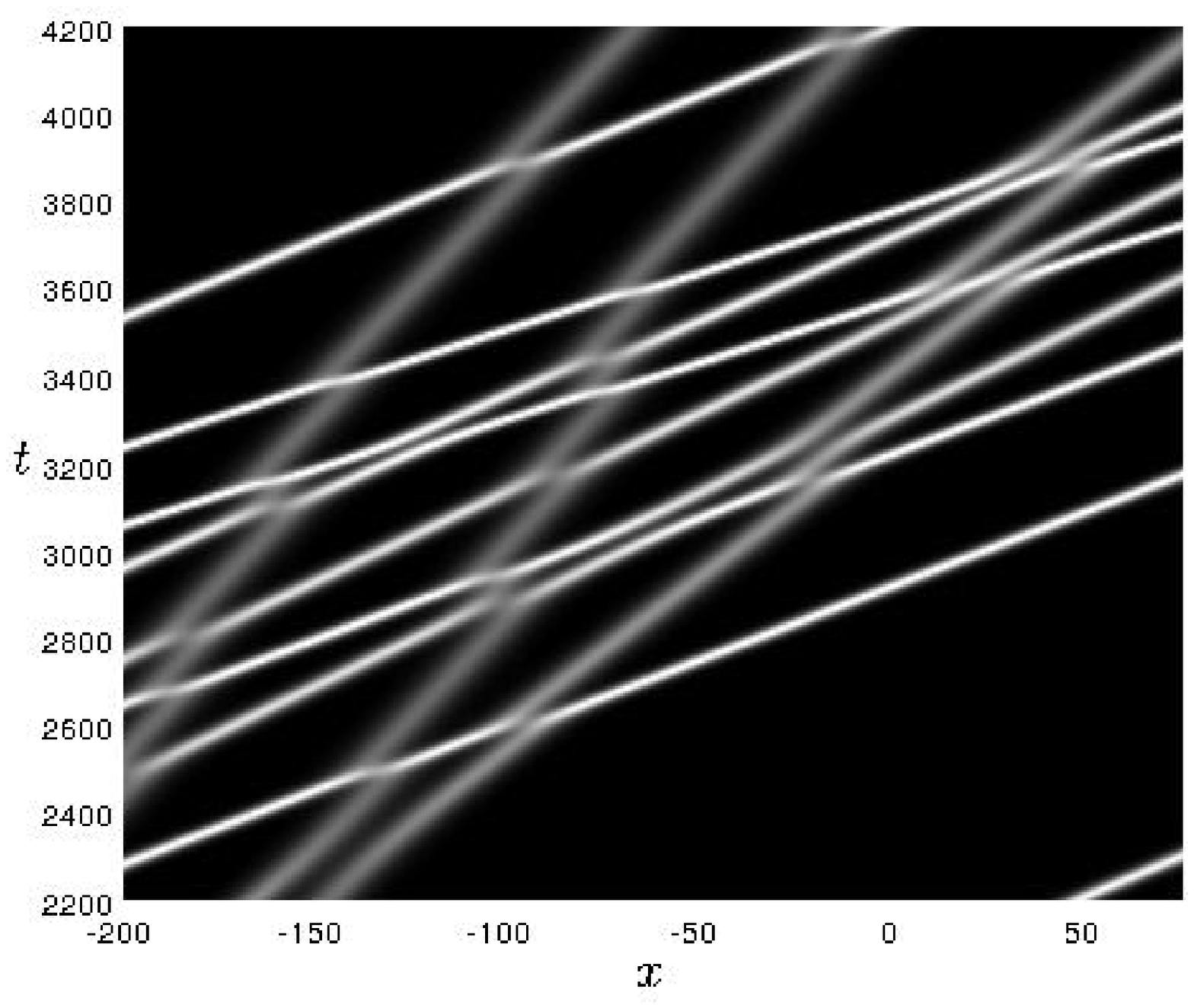}}
  \caption{Solitonic gas: (a) space-time evolution of a random ensemble of solitons under the KdV dynamics and (b) a space-time region rich of soliton collisions.}
  \label{fig:GasXT}
\end{figure}

\begin{figure}
  \centering
  \subfigure[$\I_2(t)$]%
  {\includegraphics[width=0.49\textwidth]{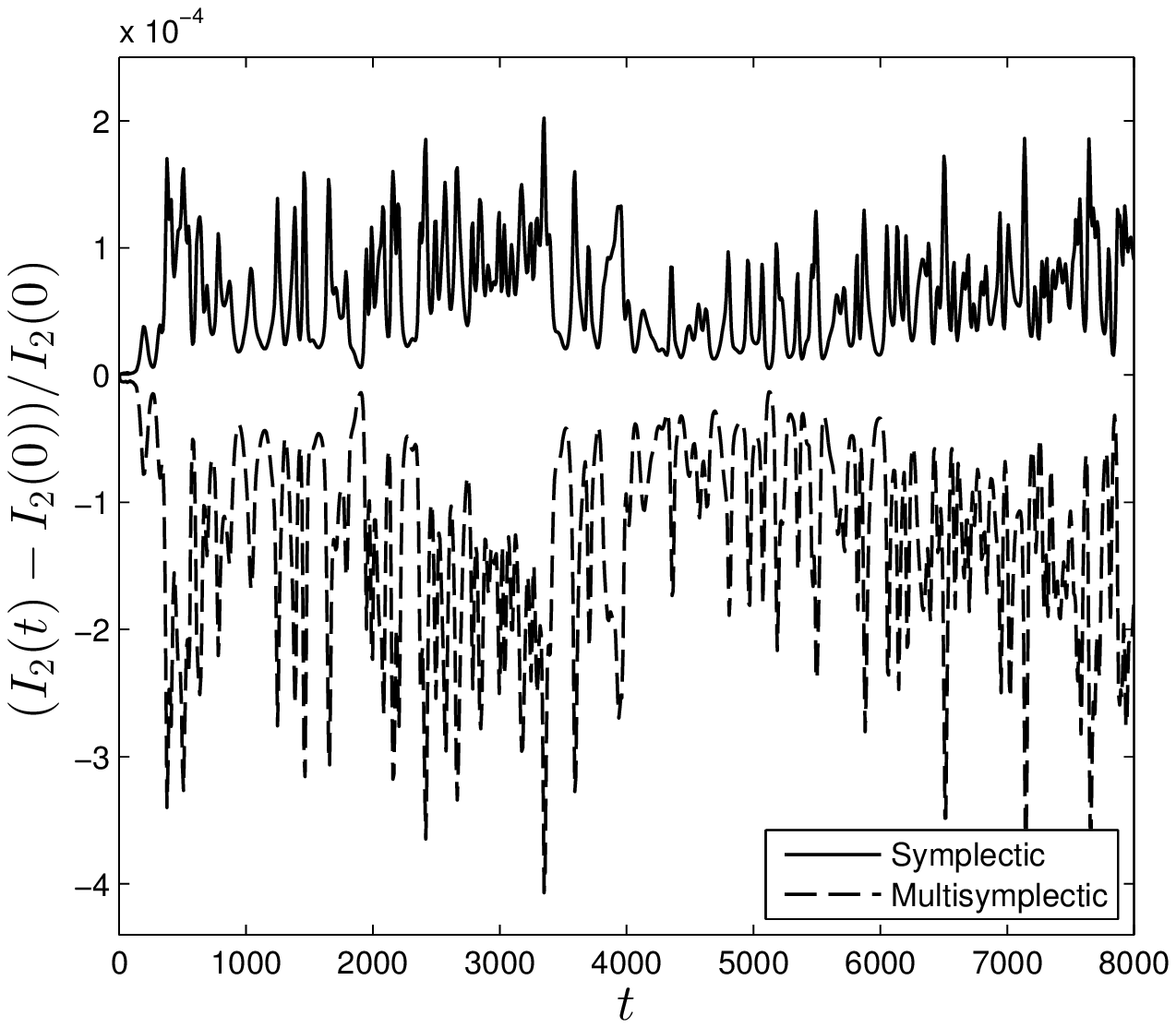}}
  \subfigure[$\I_3(t)$]%
  {\includegraphics[width=0.49\textwidth]{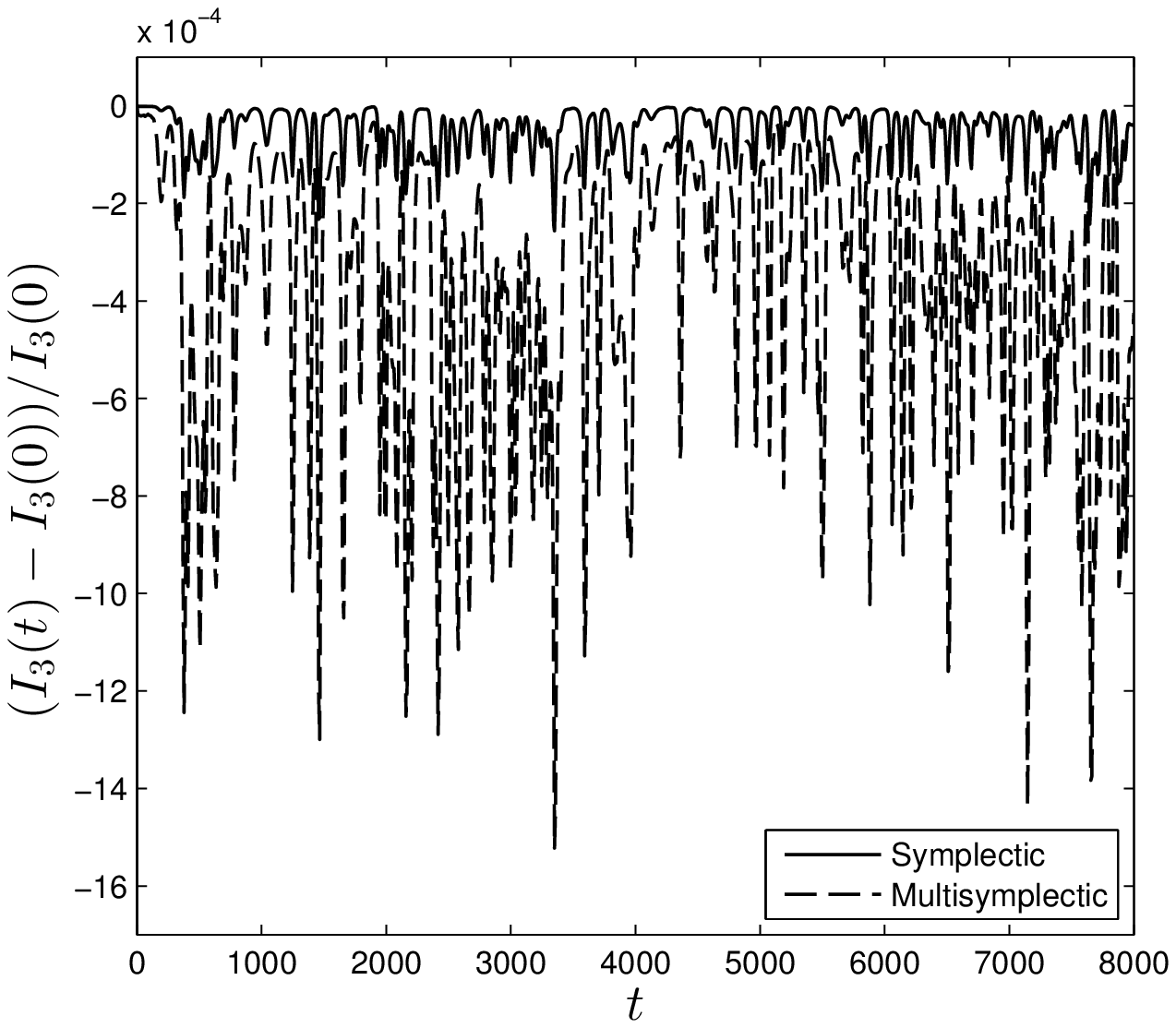}}
  \caption{Solitonic gas: time evolution of the invariants $\I_{2,3}$.}
  \label{fig:GasI23}
\end{figure}

\begin{figure}
  \centering
  \subfigure[Complete view]%
  {\includegraphics[width=0.49\textwidth]{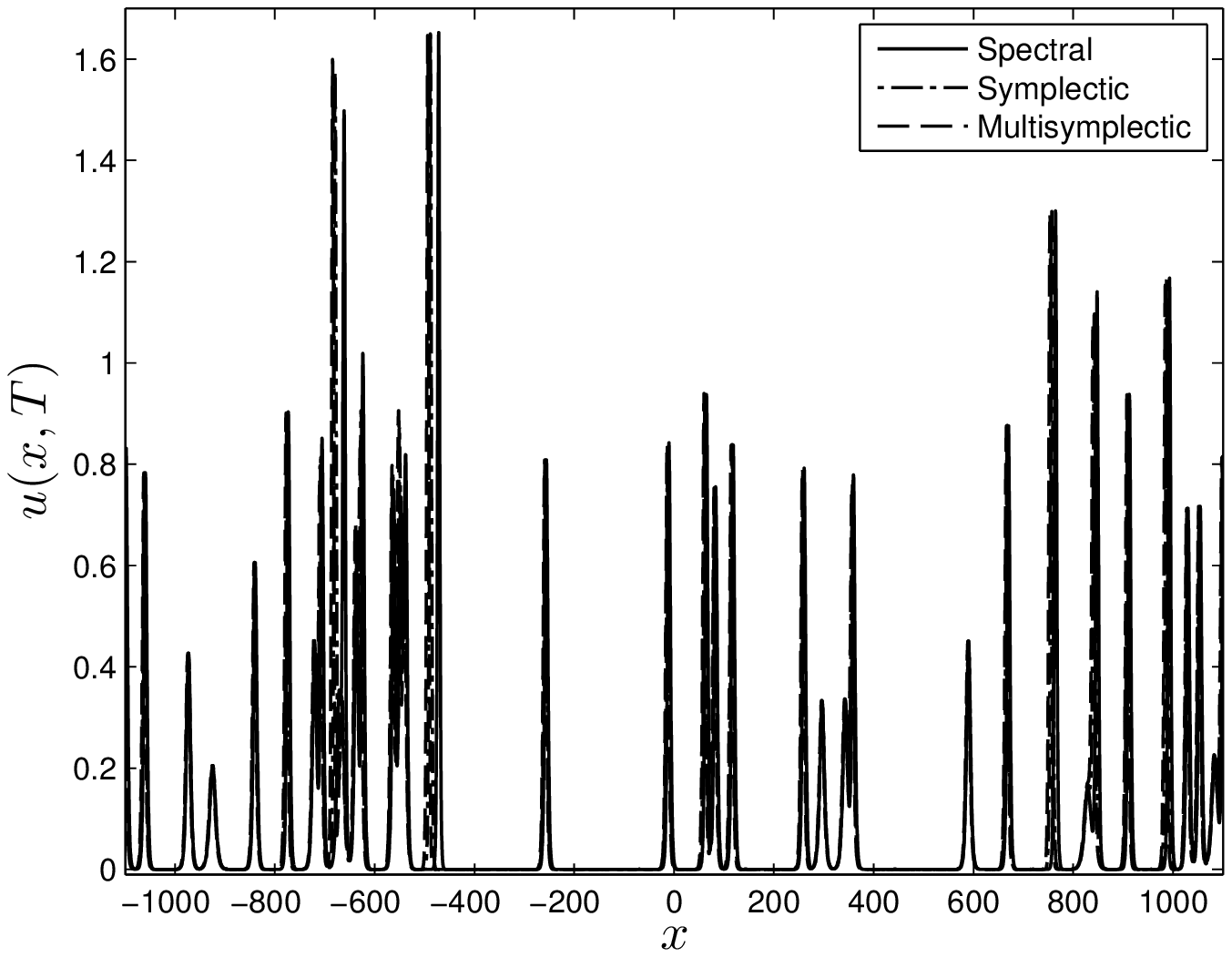}}
  \subfigure[Zoom]%
  {\includegraphics[width=0.49\textwidth]{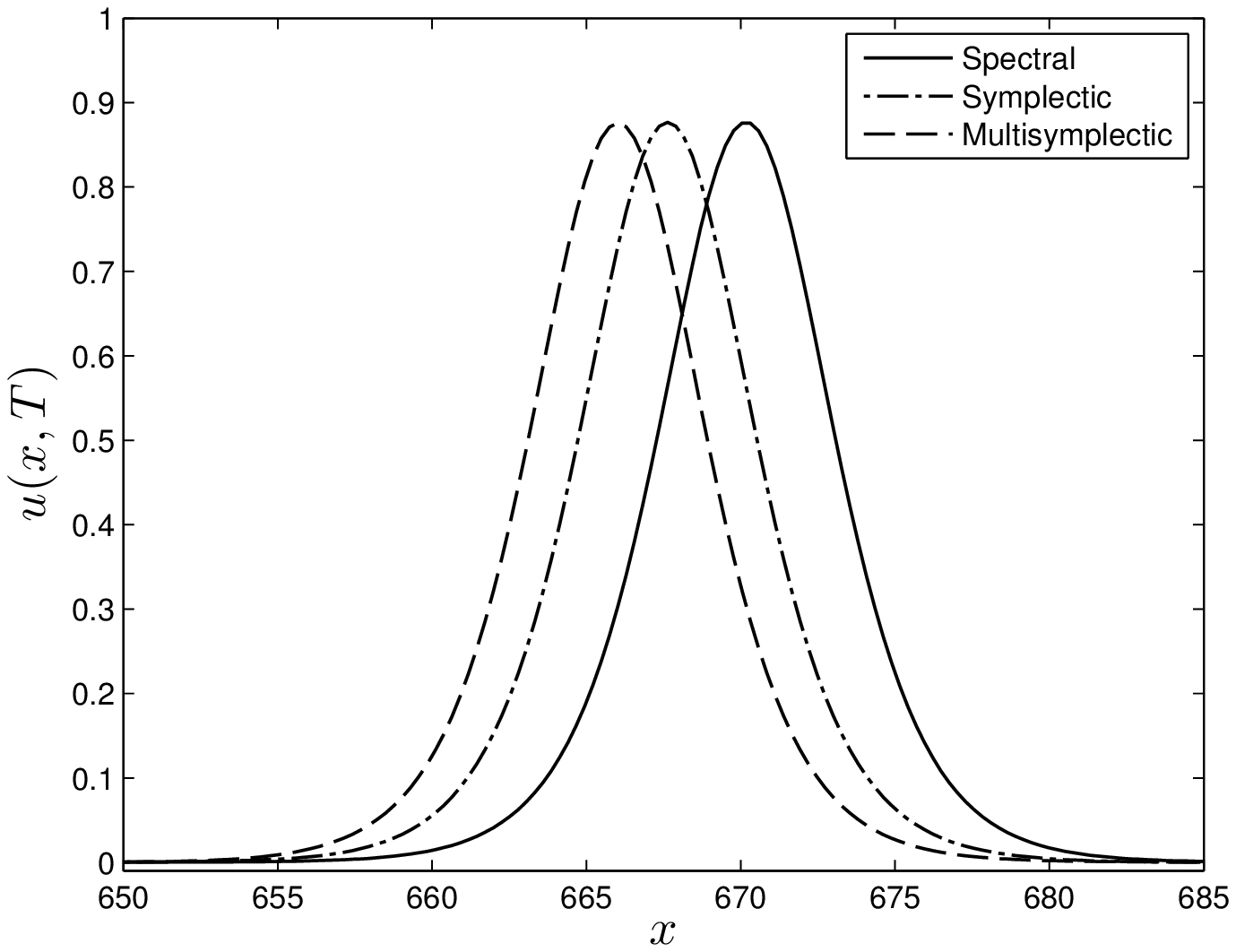}}
  \caption{Solitonic gas: (a) comparison of the numerical solutions at the final simulation time $T = 8000$ computed by the spectral, symplectic and multi-symplectic schemes, and (b) zoomed segment $650 \leq x \leq 685$.}
  \label{fig:FinalGas}
\end{figure}

The KdV dynamics is then numerically solved  up to the dimensionless time $T$ using the symplectic and multi-symplectic schemes (see  \eqref{eq:midpoint} and \eqref{eq:p8box}) as well as the spectral scheme \eqref{eq:spectral}. The final time $T$ is chosen long enough so that the tallest soliton travels twice across the entire $x$ domain (dimensionless length $\ell$). Both physical and numerical parameters are given in Table \ref{tab:gas}. As expected, solitons undergo elastic collisions as they propagate across the domain as shown in the left panel (a) of Figure \ref{fig:GasXT}, which reports the space-time evolution of the soliton gas. We observe regions in space and time rich of soliton collisions, and one of these is depicted in the right panel (b) of the same Figure. The geometric schemes behave properly solving accurately the intense interactions among solitons under a satisfactory conservation of the invariants $\I_2$ and $\I_3$, as clearly seen in Figure \ref{fig:GasI23}. We note that the symplectic scheme \eqref{eq:midpoint} performs slightly better than the multi-symplectic counterpart, showing smaller oscillations from the initial value of the invariants $\I_{2,3}$. The spectral method \eqref{eq:spectral} instead conserves both the invariants up to machine precision. Further, the left panel (a) of Figure \ref{fig:FinalGas} reports a comparison of the numerical solutions at the final time $T$ computed using the spectral, symplectic and multi-symplectic schemes, respectively. To appreciate the differences between the two geometrical solutions and the reference one from the pseudo-spectral method, the right panel (b) of the same Figure shows a zoom of the numerical wave surface in the segment $650 \leq x \leq 685$. This is a soliton whose amplitude and shape are correctly reproduced by the two geometric schemes, but a phase shift occurs due to the accumulation of local errors over the long simulation time $T$. However, the symplectic scheme provides a solution closer to the pseudo-spectral reference one. Such results and accurary are more than satisfactory, if one considers that the adopted geometric schemes are second-order accurate (see Figure \ref{fig:Err2Inf} and Table \ref{tab:rates}). Nonetheless, the phase shift error can be further reduced by decreasing the CFL number. Moreover, since all solitons experience a similar phase-shift, the whole numerical wave surface is just shifted with respect to the reference solution, thus conserving the relative position of the solitons. We conclude that geometric methods are suitable for the numerical simulation of the long-time KdV dynamics, with the symplectic scheme that performs slightly better than the multi-symplectic counterpart.

\begin{table}
  \centering
  \begin{tabular}{c|c}
    \hline\hline
    \textit{Parameter} & \textit{Value} \\
    \hline
    Domain half-length, $\ell$, $[-\ell, \ell]$ & $1100.0$ \\
    Number of discretization points, $N$ & $8192$ \\
    Discretization step, $\Delta x$ & $0.268$ \\
    CFL number & $2.0$ \\
    Time step, $\Delta t$ & $0.537$ \\
    Final simulation time, $T$ & $8000.0$ \\
    Initial gap between solitons, $\delta x$ & $60.0$ \\
    Initial shift, $x_0$ & $-\ell + 20.0$ \\
    Average solitons speed, $c_0$ & $0.25$ \\
    Variance, $\sigma^2$ & $0.0121$ \\
    Number of solitons, $M$ & $35$ \\
    \hline\hline
  \end{tabular}
  \vspace{1em}
  \caption{Solitonic gas: numerical and physical parameters used in the simulations.}
  \label{tab:gas}
\end{table}

\subsection{Statistics of a KdV random wave field}

Hereafter, we will study the evolution of a random wave field under the KdV dynamics and the associated statistics. The initial wave surface $u$ is Gaussian and given as a Fourier series of $N/2$ harmonic components ($N$ is the number of grid points) \cite{Longuet-Higgins1963, Boccotti2000}, viz.
\begin{equation*}
  u(x,0) = \sum_{j=1}^{N/2}R_j\sqrt{2S(k_j)\Delta k}\cos(k_j x + \phi_j), \quad k_j = j\Delta k, \quad \Delta k = \frac{\pi}{\ell}.
\end{equation*}

\begin{figure}
  \centering
  \includegraphics[width=0.99\textwidth]{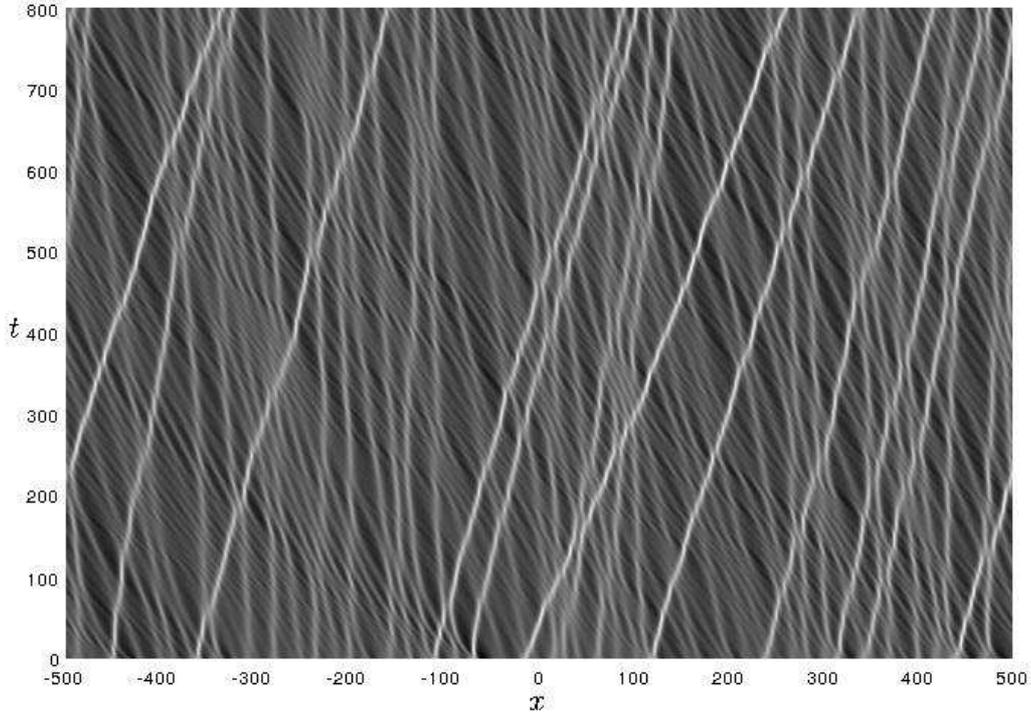}
  \caption{Random wave field: space-time evolution computed using the pseudo-spectral method (the initial wave surface is set as Gaussian).}
  \label{fig:RandXT}
\end{figure}

\begin{figure}
  \centering
  \subfigure[Complete view]%
  {\includegraphics[width=0.49\textwidth]{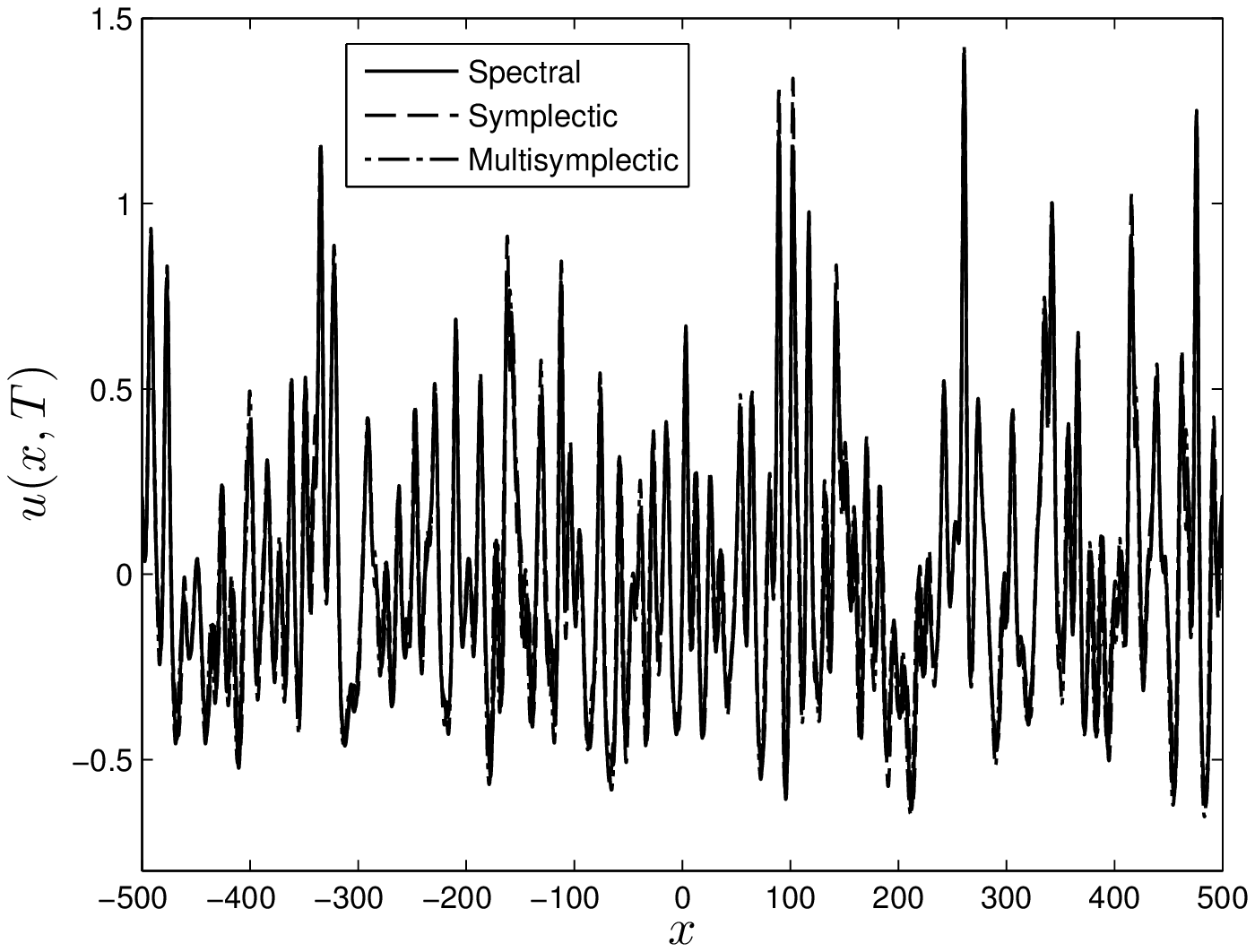}}
  \subfigure[Zoom]%
  {\includegraphics[width=0.49\textwidth]{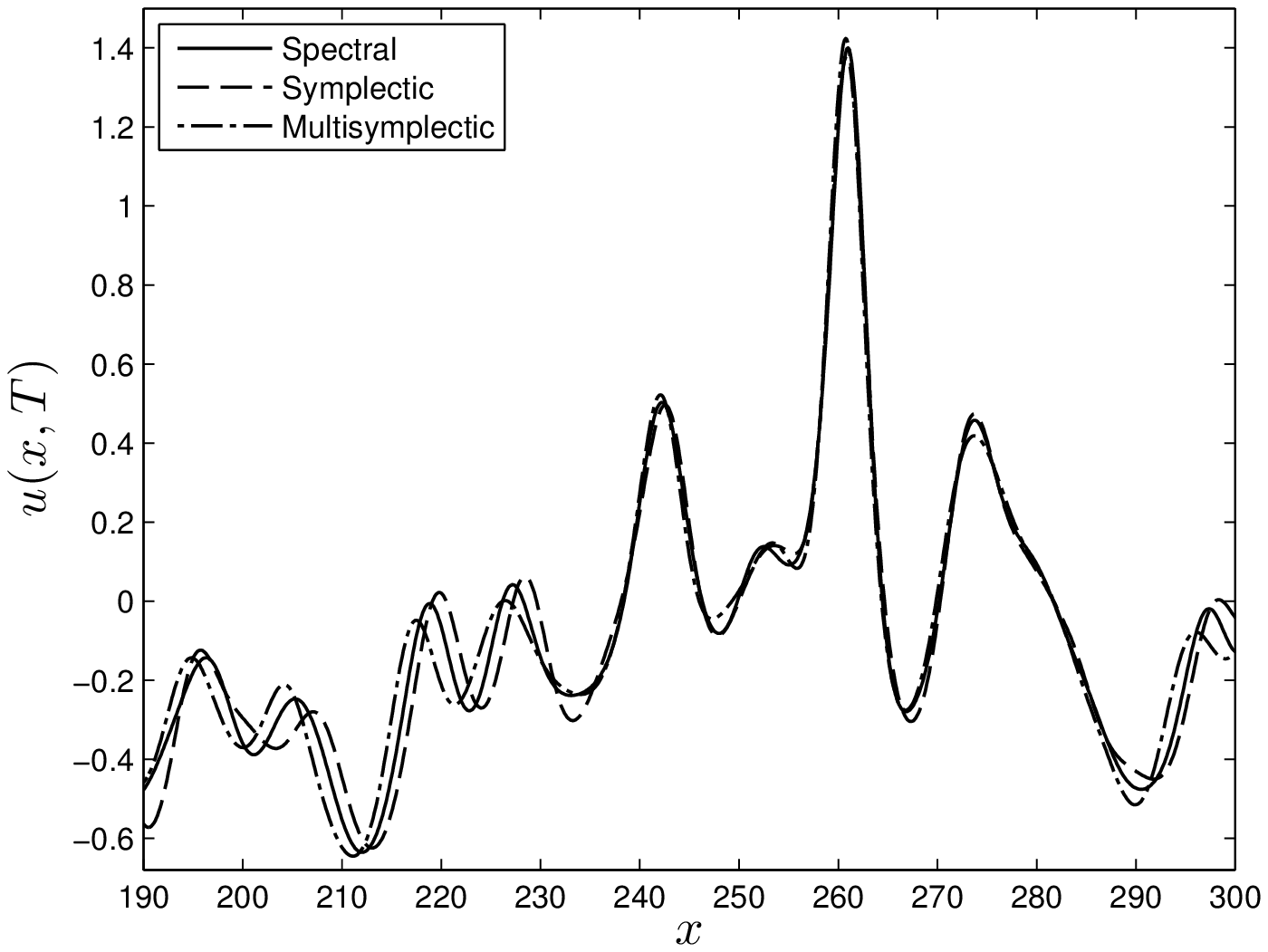}}
  \caption{Random wave field: Comparison of numerical solutions at final time $T = 800$ computed by the spectral, symplectic and multi-symplectic schemes. The right image (b) shows a zoom on the segment $190 \leq x \leq 300$.}
  \label{fig:FinalRand}
\end{figure}

\begin{figure}
  \centering
  \subfigure[Absolute values]%
  {\includegraphics[width=0.49\textwidth]{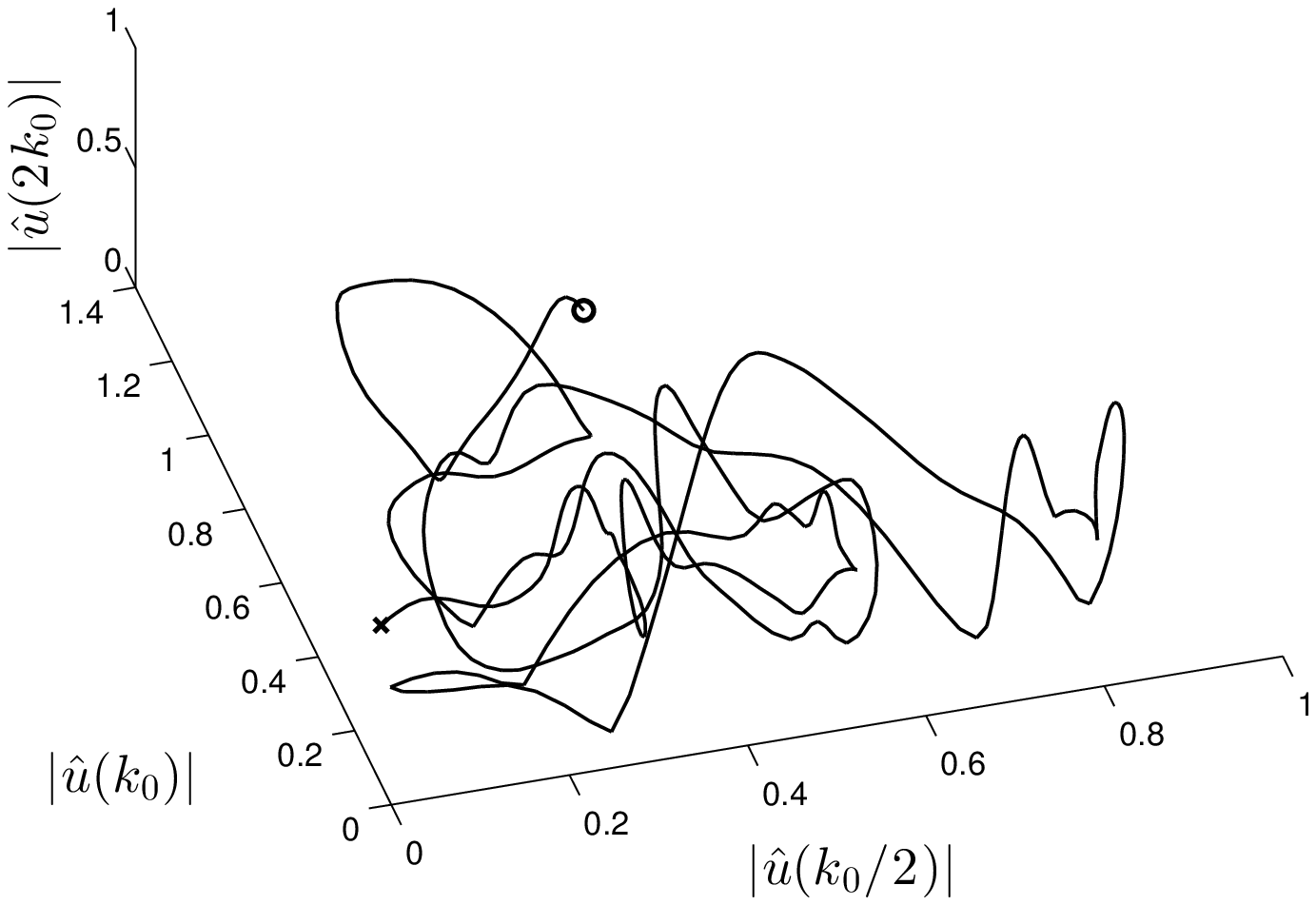}}
  \subfigure[Real parts]%
  {\includegraphics[width=0.49\textwidth]{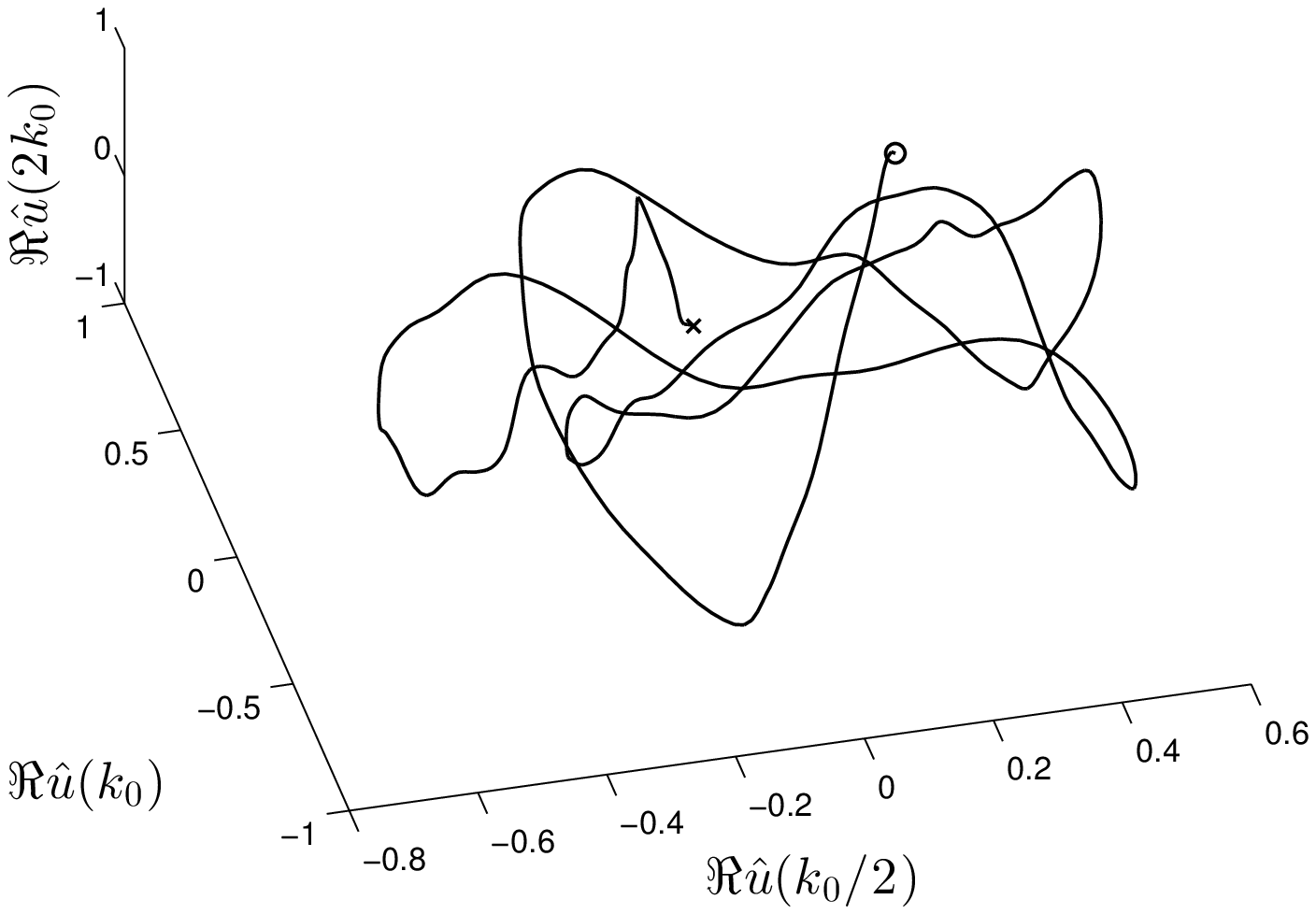}}
  \caption{Random wave field: time evolution of the Fourier amplitudes $\uh(k)$ projected onto a reduced subspace $\F_s$: (a) $\F_s = \Span\bigl\{|\uh(k_0/2)|, |\uh(k_0)|, |\uh(2k_0)|\bigr\}$, and (b) $\F_s=\Span\bigl\{\Re\uh(k_0/2), \Re\uh(k_0), \Re\uh(2k_0)\bigr\}$. The starting point of the trajectory is denoted by $\circ$ and the end by $\times$.}
  \label{fig:Orbits}
\end{figure}

\begin{figure}
  \centering
  \subfigure[$A(t)$]%
  {\includegraphics[width=0.49\textwidth]{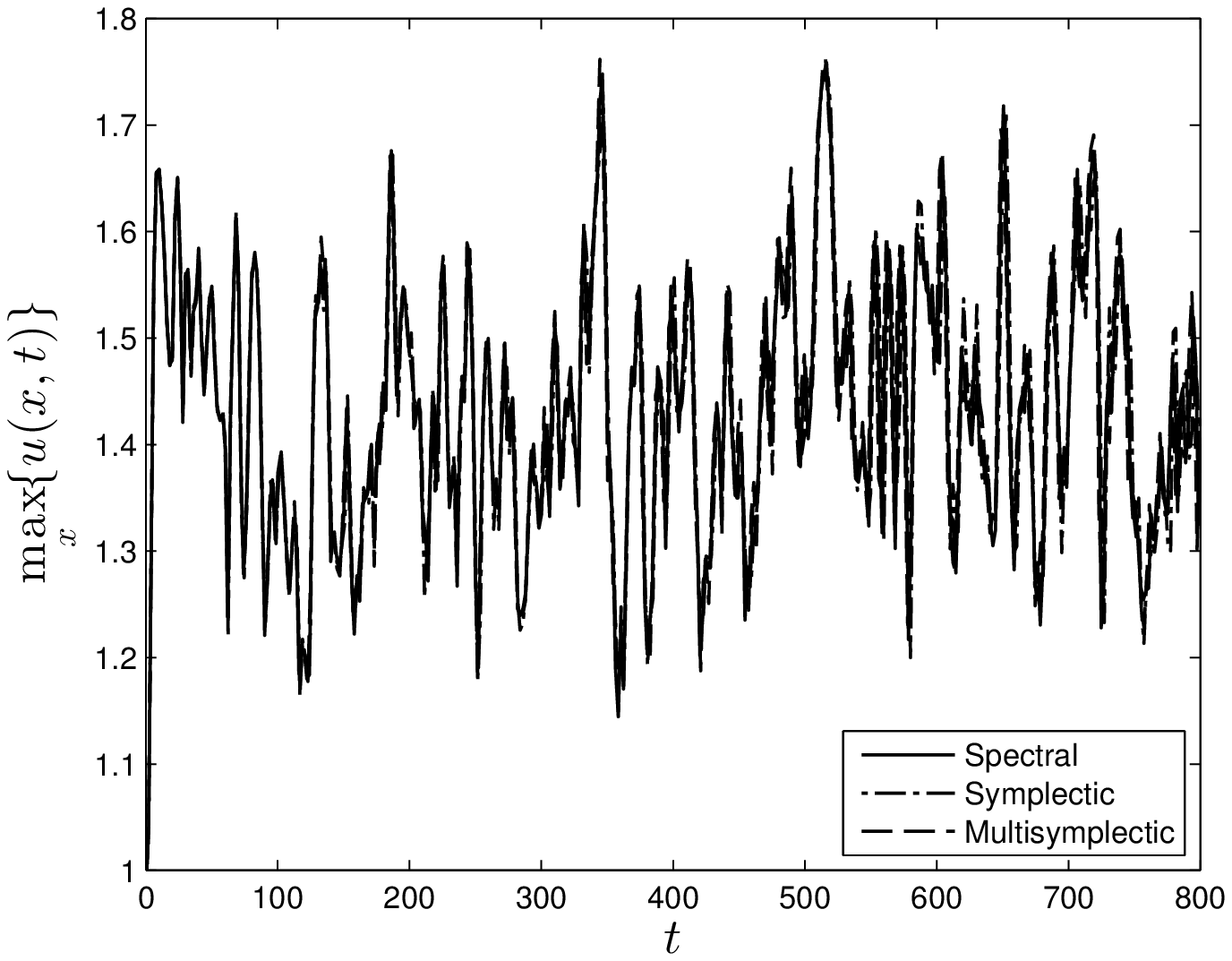}}
  \subfigure[Zoom]%
  {\includegraphics[width=0.49\textwidth]{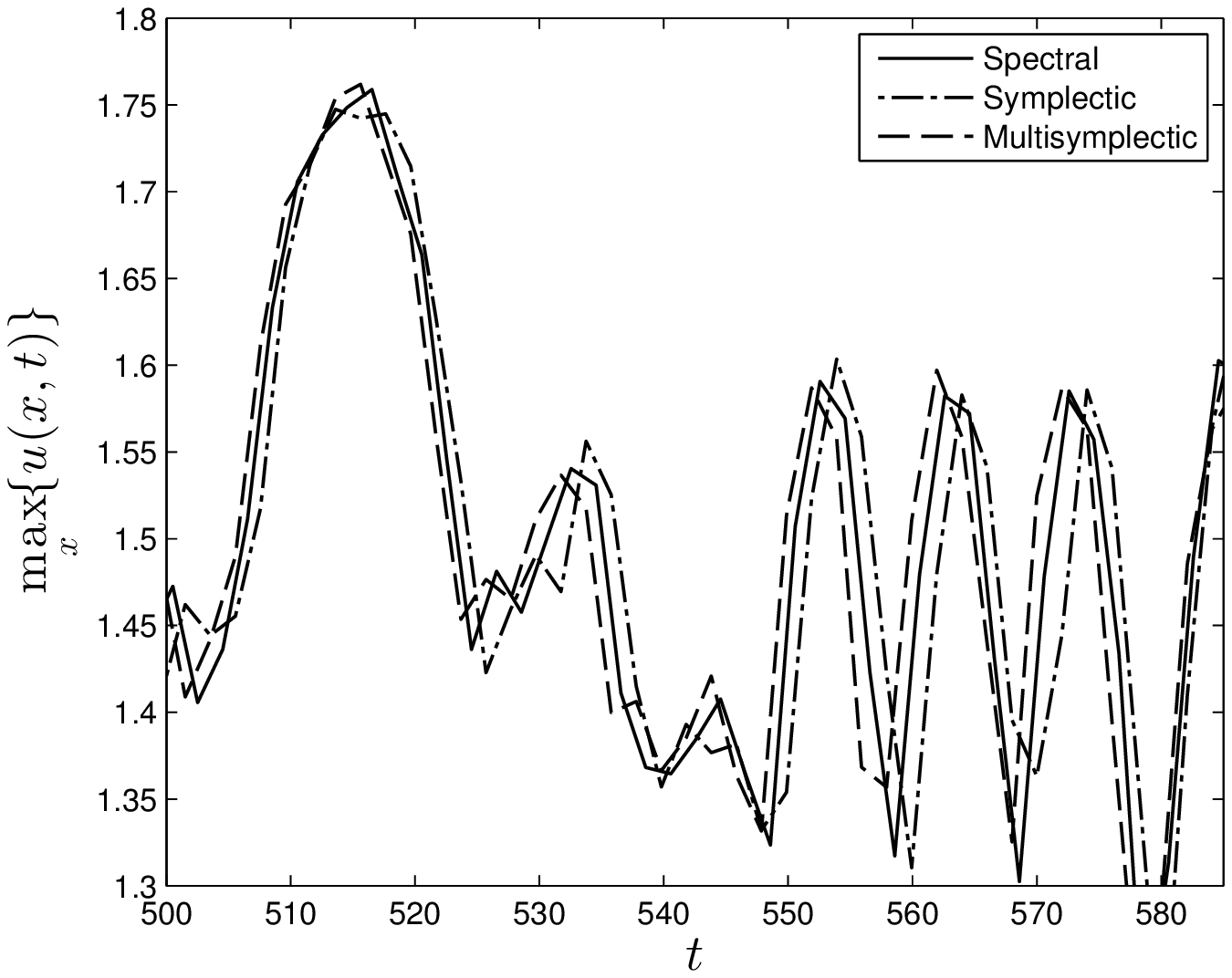}}
  \caption{Random wave field: Maximum instantaneous amplitude $A(t)$ observed during the evolution of an initial Gaussian wave field. The right panel (b) shows a zoom for $500 \leq t \leq 585$.}
  \label{fig:MaxAmp}
\end{figure}

\begin{figure}
  \centering
  \subfigure[Excess kurtosis]%
  {\includegraphics[width=0.49\textwidth]{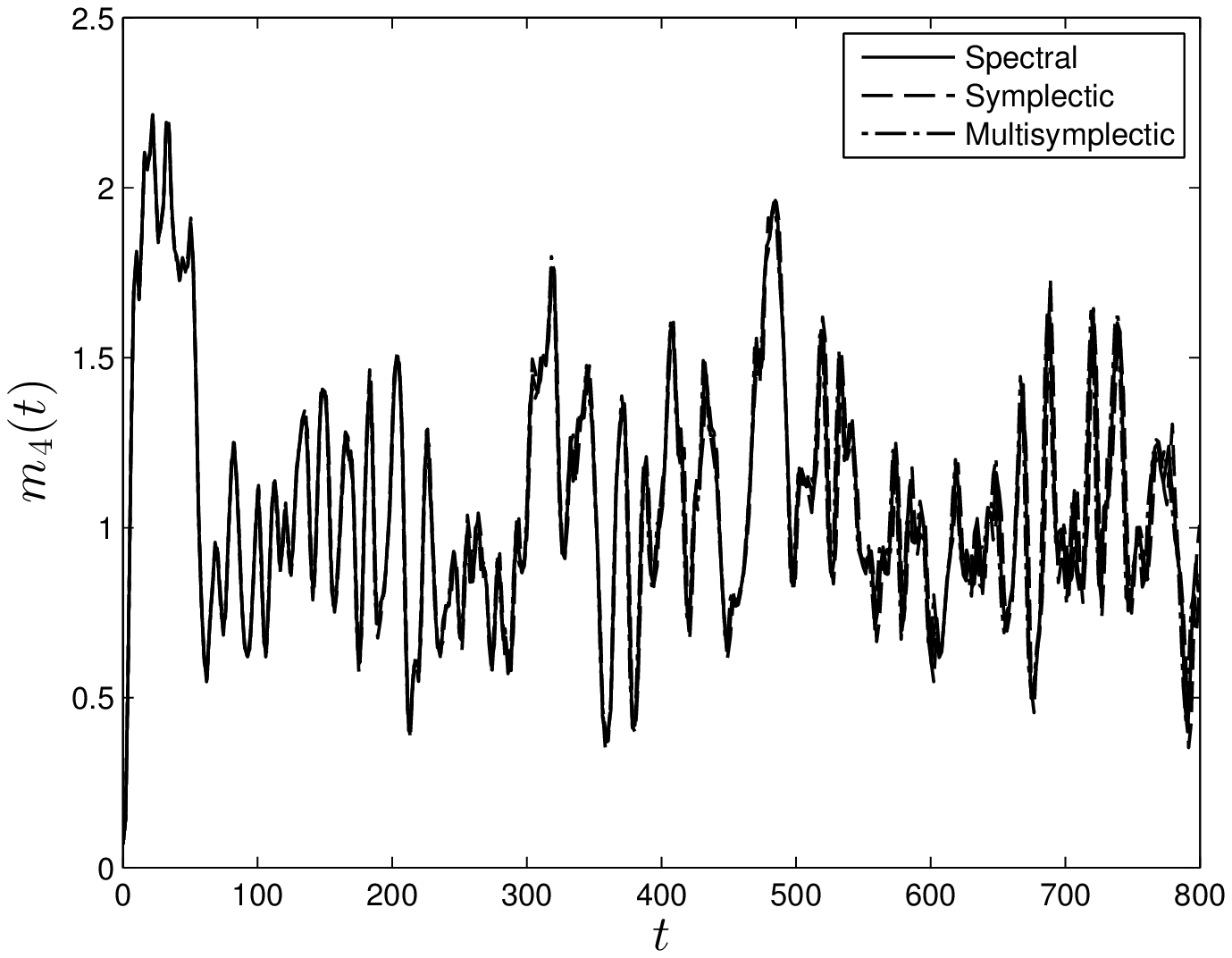}}
  \subfigure[Skewness]%
  {\includegraphics[width=0.49\textwidth]{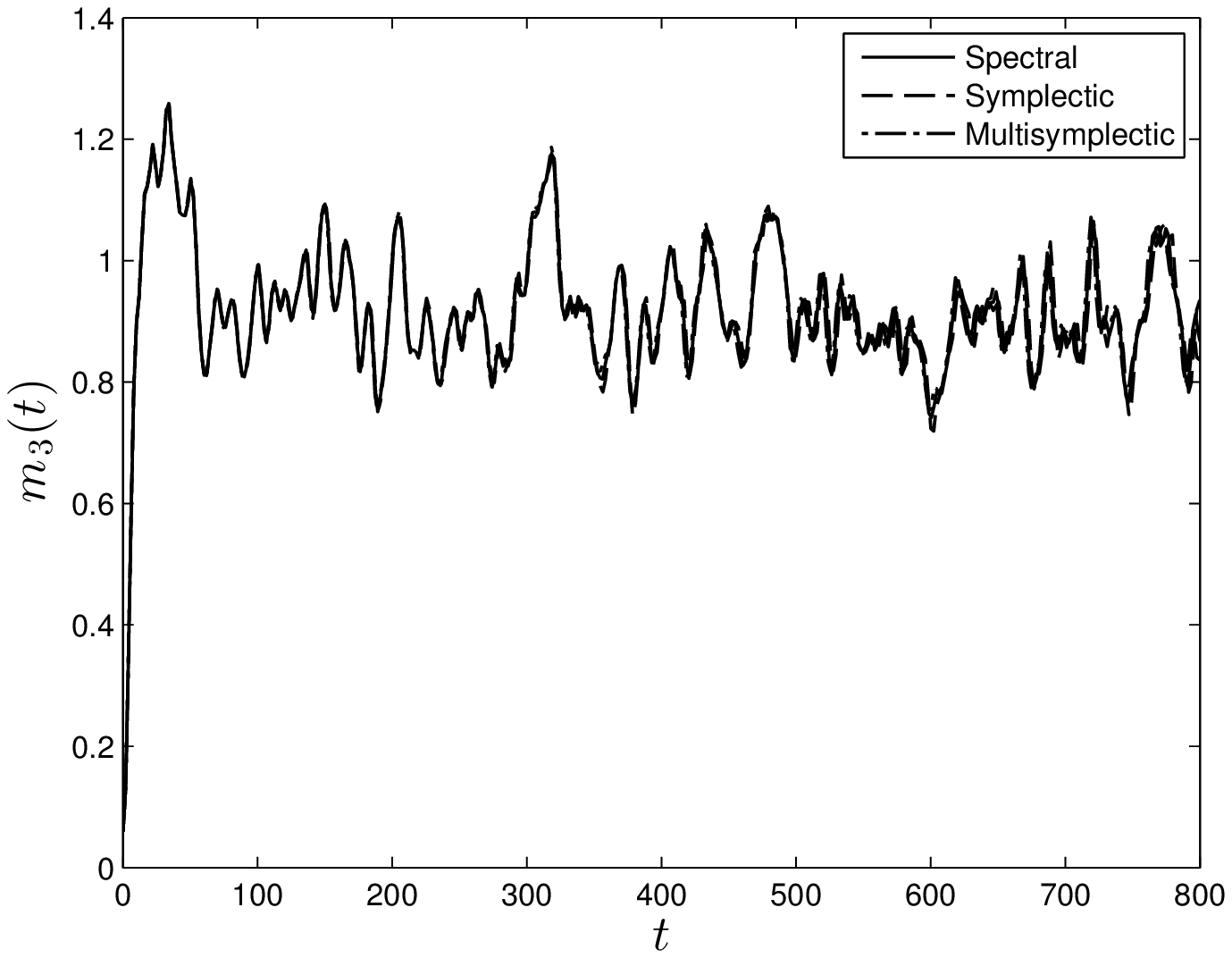}}
  \caption{Random wave field: Evolution of the excess kurtosis (a) and skewness (b) of an initial Gaussian field under the KdV dynamics.}
  \label{fig:KurtSkew}
\end{figure}

\begin{figure}
  \centering
  \subfigure[ ]%
  {\includegraphics[width=0.49\textwidth]{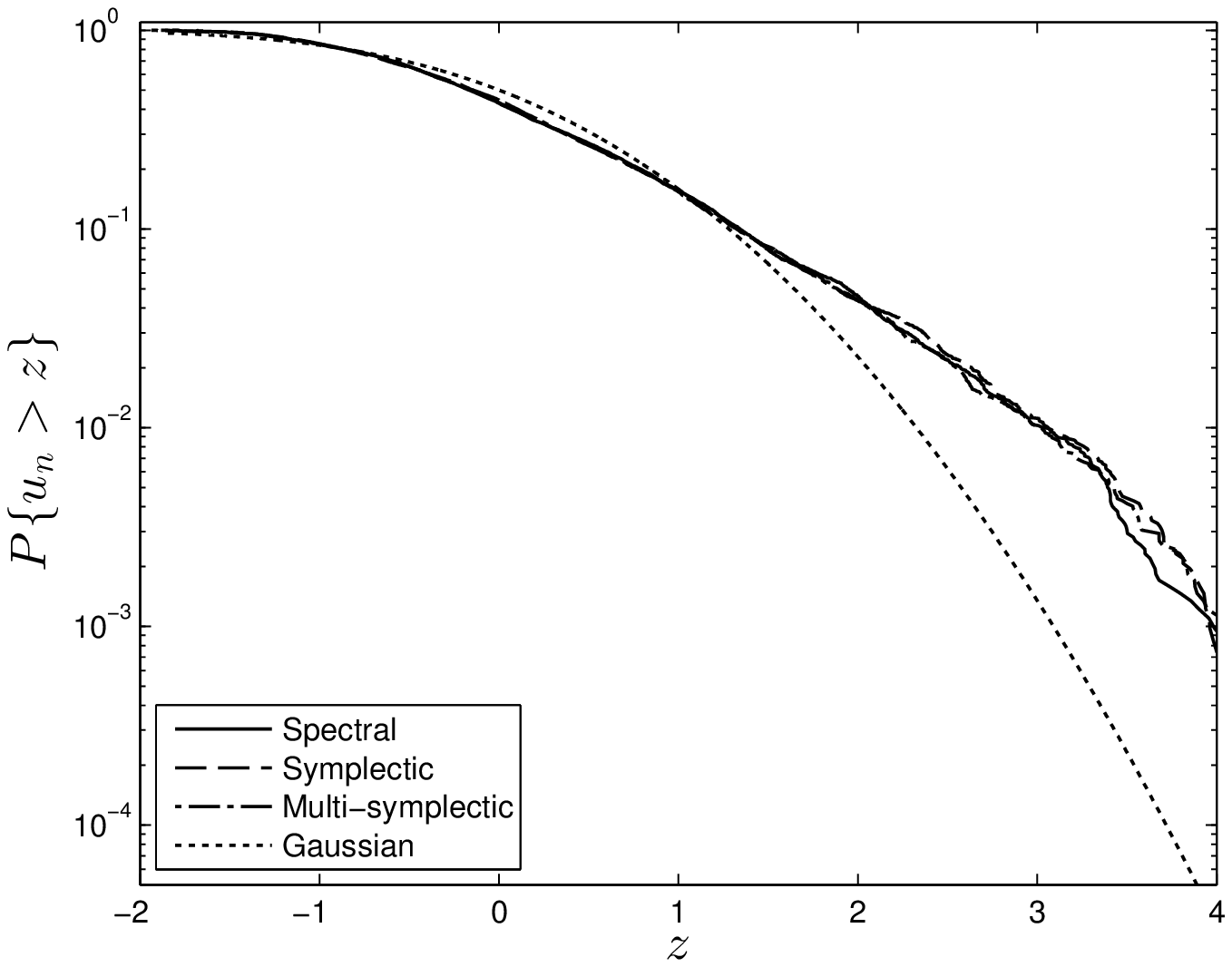}}
  \subfigure[ ]%
  {\includegraphics[width=0.49\textwidth]{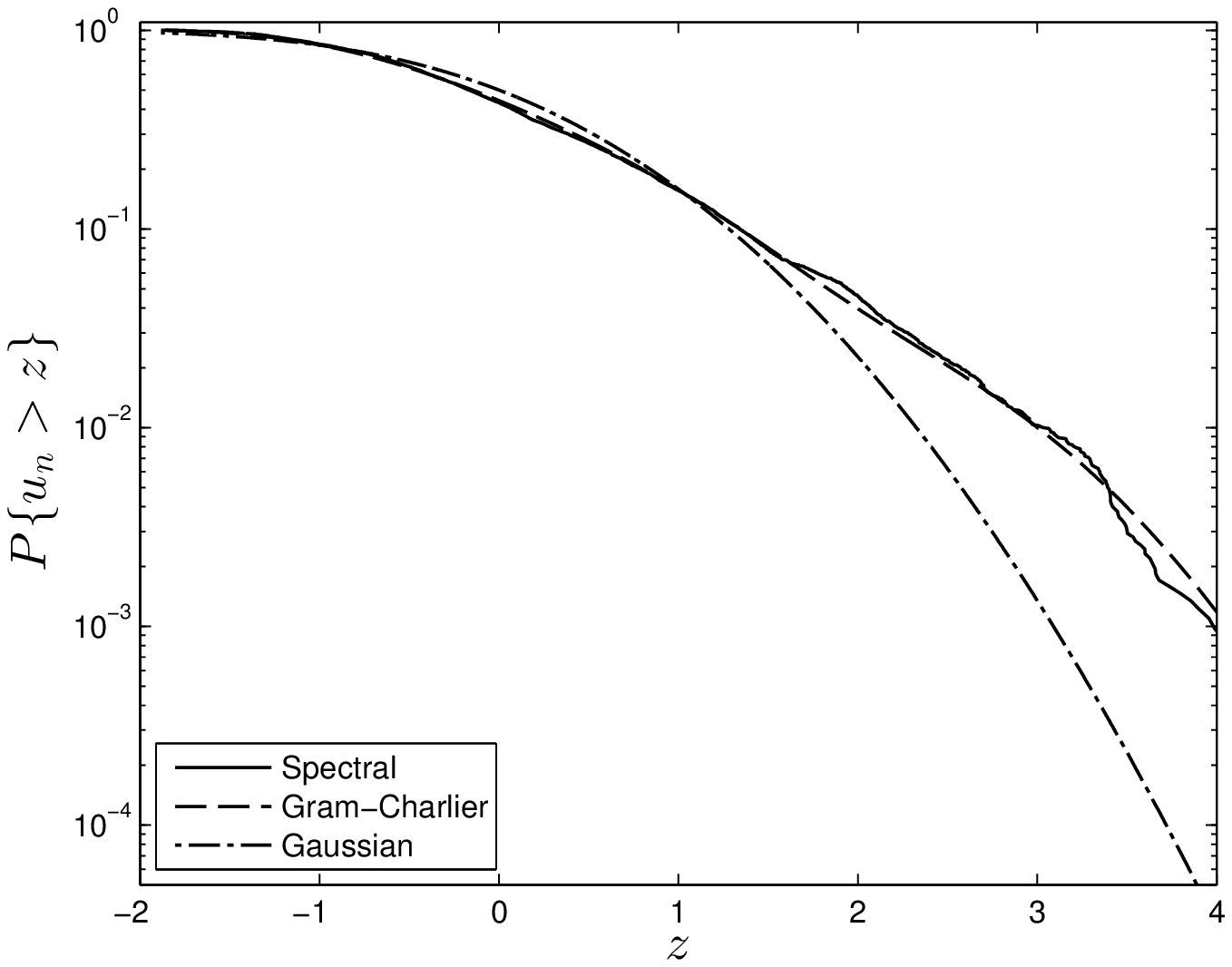}}
  \caption{Random wave field: (a) probability of exceedance of the normalized wave surface $u_n$  predicted by the numerical schemesm, and (b) observed statistics against Gaussian and Gram-Charlier distributions.}
  \label{fig:Prob}
\end{figure}

Here, 
\begin{equation*}
  S(k) = \frac{1}{\sqrt{2\pi\sigma^2}} e^{-\frac{(k-k_0)^2}{2\sigma^2}},
\end{equation*}
with $\sigma^2$ as the variance of the wave surface (see \cite{Longuet-Higgins1963}), the amplitudes
\begin{equation*}
  R_j = \sqrt{\xi_1^2 + \xi_2^2}, \qquad \xi_{1,2} \sim \N(0,1), 
\end{equation*}
are Rayleigh-distrubuted and the phases $\phi_j$'s are random and uniformly distributed in $(0, 2\pi)$. The space-time evolution resolved using the pseudo-spectral scheme is reported in Figure \ref{fig:RandXT}, which shows a gas of several solitons interacting over a dispersive wave background. Figure \ref{fig:FinalRand} reports also the comparison of the numerical solutions at the final simulation time $T$ computed by the spectral, symplectic and multi-symplectic schemes, respectively. The agreement of the two variational numerical solutions with the reference spectral one is excellent. The dynamics in the Fourier space $\F(k)$ is interesting since it shows how energy is exchanged by the harmonic components $\uh(k)$ due to nonlinear interactions. This is clearly seen in the left panel (a) of Figure \ref{fig:Orbits}, which shows the Fourier projection onto the subspace $\F_s = \Span\bigl\{|\uh(k_0/2)|, |\uh(k_0)|, |\uh(2k_0)|\bigr\}$, where $k_0$ is the dominant wavenumber. Note that the solution trajectory ergodically visits many regions of $\F_s$ and it is expected to cover the entire subspace in the long time. In the right panel (b) of the same Figure, we also report the projection on the subspace $\F_s = \Span\bigl\{\Re\uh(k_0/2), \Re\uh(k_0), \Re\uh(2k_0)\bigr\}$, with $\Re\uh$ denoting the real part of $\uh$. The maximum instantaneous amplitude $A(t) := \max\limits_{-\ell\leq x \leq\ell}\{u(x,t)\}$ of the wave field over the space domain is shown in Figure \ref{fig:MaxAmp}. Note the good agreement of the symplectic solutions with the reference spectral one (see also panel (b) in the same figure, which depicts the amplitude evolution during $500 \leq t \leq 585$). Figure \ref{fig:KurtSkew} reports the time evolution of the excess kurtosis $m_4 := \langle u_n^4\rangle/m_2^2 - 3$ (left panel (a)) and skewness $m_3 = \langle u_n^3\rangle/m_2^{3/2}$ (right panel (b)), where $u_n := (u - \langle u\rangle)/m_2^{1/2}$ is the normalized wave surface and $m_2 := \langle u_n^2\rangle$ is the variance of $u$ with respect to the mean $\langle u\rangle$. Clearly, under the KdV dynamics the wave field deviates statistically from the initial Gaussian conditions reaching an ergodic non-Gaussian steady state. The large excess kurtosis is due to the interaction among solitons, which yields the sudden arise of large peaks from the wave background intermittently (see, for example, right panel of Figure \ref{fig:FinalRand}). As a result, at steady state the probability of the exceedance $\Pr\{u_n > z\}$ from Gaussian and it is well represented by the truncated Gram--Charlier form \cite{Longuet-Higgins1963, Tayfun2007, Fedele2008a}
\begin{equation*}
  \Pr\{u_n > z\} = \frac12\erfc\bigl(\frac{z}{\sqrt{2}}\bigr) + \frac{1}{2\sqrt{2\pi}}e^{-\frac{z^2}{2}}\Bigl(\frac{m_3}{3}(z^2-1) + \frac{m_4}{12}z(z^2-3) + \frac{m_3^2}{36}z(z^4 - 10z^2 + 15)\Bigr).
\end{equation*}
as clearly shown in the right panel (b) of Figure \ref{fig:Prob}. Finally, note that the solutions of the three adopted numerical schemes yield the same statistical distributions at the final simulation time $T$ (see left panel (a) of the same Figure).

\begin{table}
  \centering
  \begin{tabular}{c|c}
    \hline\hline
    \textit{Parameter} & \textit{Value} \\
    \hline
    Domain half-length, $\ell$, $[-\ell, \ell]$ & $500.0$ \\
    Number of discretization points, $N$ & $8192$ \\
    Discretization step, $\Delta x$ & $0.122$ \\
    CFL number & $0.75$ \\
    Time step, $\Delta t$ & $0.0916$ \\
    Final simulation time, $T$ & $800.0$ \\
    Peak wavenumber, $k_0$ & $0.2$ \\
    Variance, $\sigma^2$ & $0.0225$ \\
    \hline\hline
  \end{tabular}
  \vspace{1em}
  \caption{Random wave field: numerical and physical parameters used in the simulations.}
  \label{tab:rand}
\end{table}

\section{Discussion and conclusions}\label{sec:concl}

In this study we tried to assess the realistic performance of some nowadays popular geometric discretizations for PDEs. Namely, we considered symplectic and multi-symplectic schemes of the same order of accuracy. As the main PDE we opted for the celebrated Korteweg -- de Vries (KdV) equation to model nonlinear dispersive waves. This equation arises in various fields of physics and engineering and it has been used in several recent studies of solitonic and wave turbulence \cite{Salupere2003, Pelinovsky2006, Zahibo2009, Sergeeva2011}. It possesses both Hamiltonian and multi-symplectic structures, which provide the basis for the formulation of promising geometric schemes \cite{Ascher2004, Ascher2005} for modeling complex wave phenomena such as wave turbulence (see, for example \cite{Cai2001}). However, to date these schemes have only been applied to academic test-cases. We thus considered several test-cases in order to assess the ability of geometric discretizations to simulate the evolution of complex nonlinear wave processes over long integration times. As a reference solution we opted for that by a highly accurate pseudo-spectral method.

As a representative of geometric schemes, we chose the classical symplectic mid-point rule \cite{Hairer2002, Ascher2004, Leimkuhler2004} combined with a variational discretization in space. For comparison, we included also the multi-symplectic 8-point box scheme proposed by U.~\textsc{Ascher} \& R.~\textsc{McLachlan} (2004) \cite{Ascher2004}, which treats space and time on equal footing. Both schemes have formally the same order of accuracy, so to have a fair comparison between their performances. Note that in their study \textsc{Ascher} \& \textsc{McLachlan} \cite{Ascher2004} proposed also a semi-explicit symplectic scheme, which we did not consider in this work because it is prone to numerical instabilities. For a robust numerical solution of the  long-time dynamics apparently one needs fully implicit discretization schemes. Moreover, our numerical results indicate that the 8-point multi-symplectic box scheme does not bring any significant advantage over the symplectic one. This surprising conclusion is probably due to the fact that the box scheme uses a non-symmetric stencil for the discretization of the third order spatial derivative.

We have exploited symplectic schemes to study the long-time KdV dynamics of a solitonic gas and the statistics of random wave fields. Our results suggest that that the multi-symplectic, but also \emph{fully implicit} symplectic schemes are completely suitable for complex simulations of the long-time dynamics of nonlinear waves, as those needed to model wave turbulence. Not only they are able to represent correctly the averaged statistical quantities, but also they have robust convergence properties, viz. the numerical error $\eps \sim O(\Delta x^n)$. Namely, if we fix a random initial state, the numerical trajectory will not significantly deviate from the reference solution during very long evolution times. This surprising performance of symplectic and multi-symplectic schemes is certainly due to their superior ability to preserve the geometric properties in phase space. In future investigations we will use geometric schemes to study the evolution of random turbulent states in non-periodic (wall bounded or even more complex) domains to see the effects of confinement and of boundary conditions on statistical characteristics of the wave field.

\section*{Acknowledgements}

D.~Dutykh would like to acknowledge the hospitality of the Basque Center for Applied Mathematics (BCAM) during his visit in March, 2012.

\bibliography{biblio}

\begin{thebibliography}{10}

\bibitem{Abdullaev1995}
F.~Kh. Abdullaev, S.~A. Darmanyan, M.~R. Djumaev, A.~J. Majid, and M.~P.
  S\o~rensen.
\newblock {Evolution of randomly perturbed Korteweg-de Vries solitons}.
\newblock {\em Phys. Rev. E}, 52(4):3577--3583, 1995.

\bibitem{Annenkov2006}
S.~Y. Annenkov and V.~I. Shrira.
\newblock {Direct numerical simulation of downshift and inverse cascade for
  water wave turbulence.}
\newblock {\em Phys. Rev. Lett}, 96(20):204501, 2006.

\bibitem{Arnold1980}
D.~N. Arnold and R.~Winther.
\newblock {A conservative finite element method for the Korteweg-de Vries
  equation}.
\newblock {\em Mathematics of Computation}, 34(149):23--43, 1980.

\bibitem{Ascher2004}
U.~M. Ascher and R.~I. McLachlan.
\newblock {Multisymplectic box schemes and the Korteweg-de Vries equation}.
\newblock {\em Applied Numerical Mathematics}, 48(3-4):255--269, 2004.

\bibitem{Ascher2005}
U.~M. Ascher and R.~I. McLachlan.
\newblock {On Symplectic and Multisymplectic Schemes for the KdV Equation}.
\newblock {\em J. Sci. Comput.}, 25(1):83--104, 2005.

\bibitem{Benkhaldoun2008}
F.~Benkhaldoun and M.~Seaid.
\newblock {New finite-volume relaxation methods for the third-order
  differential equations}.
\newblock {\em Commun. Comput. Phys.}, 4:820--837, 2008.

\bibitem{Berger2003}
K.~M. Berger and P.~A. Milewski.
\newblock {Simulation of wave interactions and turbulence in one-dimensional
  water waves}.
\newblock {\em SIAM Journal on Applied Mathematics}, 63(4):1121--1140, 2003.

\bibitem{Boccotti2000}
P.~Boccotti.
\newblock {\em {Wave Mechanics for Ocean Engineering}}.
\newblock Elsevier Sciences, Oxford, 2000.

\bibitem{Bona2007}
J.~L. Bona, V.~A. Dougalis, and D.~E. Mitsotakis.
\newblock {Numerical solution of KdV-KdV systems of Boussinesq equations: I.
  The numerical scheme and generalized solitary waves}.
\newblock {\em Mat. Comp. Simul.}, 74:214--228, 2007.

\bibitem{Boyd2000}
J.~P. Boyd.
\newblock {\em {Chebyshev and Fourier Spectral Methods}}.
\newblock 2nd edition, 2000.

\bibitem{Bridges2001}
T.~J. Bridges and S.~Reich.
\newblock {Multi-symplectic integrators: numerical schemes for Hamiltonian PDEs
  that conserve symplecticity}.
\newblock {\em Physics Letters A}, 284(4-5):184--193, 2001.

\bibitem{Bridges2006}
T.~J. Bridges and S.~Reich.
\newblock {Numerical methods for Hamiltonian PDEs}.
\newblock {\em J. Phys. A: Math. Gen}, 39:5287--5320, 2006.

\bibitem{Cai2001}
D.~Cai, A.~J. Majda, D.~W. McLaughlin, and E.~G. Tabak.
\newblock {Dispersive wave turbulence in one dimension}.
\newblock {\em Physica D: Nonlinear Phenomena}, 152-153(1-2):551--572, 2001.

\bibitem{Calvo1993}
M.~P. Calvo and J.-M. Sanz-Serna.
\newblock {Symplectic numerical methods for Hamiltonian problems}.
\newblock In {\em Physics Computing 92}, pages 153--160. World Scientific,
  Singapore, 1993.

\bibitem{Canuto2006}
C.~Canuto, M.~Y. Hussaini, A.~Quarteroni, and T.~A. Zang.
\newblock {\em {Spectral Methods Fundamentals in Single Domains}}.
\newblock Scientific Computation. Springer-Verlag Berlin Heidelberg, 2006.

\bibitem{Chhay2011}
M.~Chhay, E.~Hoarau, A.~Hamdouni, and P.~Sagaut.
\newblock {Comparison of some Lie-symmetry-based integrators}.
\newblock {\em J. Comp. Phys.}, 230(5):2174--2188, 2011.

\bibitem{Courant1928}
R.~Courant, K.~Friedrichs, and H.~Lewy.
\newblock {\"{U}ber die partiellen Differenzengleichungen der mathematischen
  Physik}.
\newblock {\em Mathematische Annalen}, 100(1):32--74, 1928.

\bibitem{Courant1967}
R.~Courant, K.~Friedrichs, and H.~Lewy.
\newblock {On the partial difference equations of mathematical physics. English
  translation of the 1928 German original}.
\newblock {\em IBM Journal}, pages 215--234, 1967.

\bibitem{DeBouard2007}
A.~{De Bouard} and A.~Debussche.
\newblock {Random modulation of solitons for the stochastic Korteweg-de Vries
  equation}.
\newblock {\em Ann. Inst. Henri Poincar\'{e}}, 24(2):251--278, 2007.

\bibitem{Debussche1999}
A.~Debussche.
\newblock {Numerical simulation of the stochastic Korteweg-de Vries equation}.
\newblock {\em Physica D: Nonlinear Phenomena}, 134(2):200--226, 1999.

\bibitem{Dingemans1997}
M.~W. Dingemans.
\newblock {\em {Water wave propagation over uneven bottom}}.
\newblock World Scientific, Singapore, 1997.

\bibitem{Dutykh2010e}
D.~Dutykh, Th. Katsaounis, and D.~Mitsotakis.
\newblock {Finite volume methods for unidirectional dispersive wave models}.
\newblock {\em Submitted}, 2010.

\bibitem{El2005a}
G.~El and A.~Kamchatnov.
\newblock {Kinetic Equation for a Dense Soliton Gas}.
\newblock {\em Phys. Rev. Lett}, 95(20):204101, November 2005.

\bibitem{Fedele2008a}
F.~Fedele.
\newblock {Rogue wave in oceanic turbulence}.
\newblock {\em Physica D}, 237:2127--2131, 2008.

\bibitem{FFTW98}
M.~Frigo and S.~G. Johnson.
\newblock {FFTW: An adaptive software architecture for the FFT}.
\newblock In {\em Proc. 1998 IEEE Intl. Conf. Acoustics Speech and Signal
  Processing}, volume~3, pages 1381--1384. IEEE, 1998.

\bibitem{Frigo2005}
M.~Frigo and S.~G. Johnson.
\newblock {The Design and Implementation of FFTW3}.
\newblock {\em Proceedings of the IEEE}, 93(2):216--231, 2005.

\bibitem{Garnier2001}
J.~Garnier.
\newblock {Long-time dynamics of Korteweg-de Vries solitons driven by random
  perturbations}.
\newblock {\em Journal of Statistical Physics}, 105(5-6):789--833, 2001.

\bibitem{Hairer2002}
E.~Hairer, C.~Lubich, and G.~Wanner.
\newblock {\em {Geometric Numerical Integration}}, volume~31 of {\em Spring
  Series in Computational Mathematics}.
\newblock Springer-Verlag, Berlin, Heidelberg, second edition, 2002.

\bibitem{Helal2001}
M.~A. Helal.
\newblock {A Chebyshev spectral method for solving Korteweg-de Vries equation
  with hydrodynamical application}.
\newblock {\em Chaos, Solitons \& Fractals}, 12(5):943--950, 2001.

\bibitem{Hydon2005}
P.~E. Hydon.
\newblock {Multisymplectic conservation laws for differential and
  differential-difference equations}.
\newblock {\em Proc. R. Soc. A}, 461:1627--1637, 2005.

\bibitem{Isaacson1966}
E.~Isaacson and H.~B. Keller.
\newblock {\em {Analysis of Numerical Methods}}.
\newblock Dover Publications, 1966.

\bibitem{Islas2005}
A.~L. Islas and C.~M. Schober.
\newblock {Backward error analysis for multisymplectic discretizations of
  Hamiltonian PDEs}.
\newblock {\em Math. Comp. Simul.}, 69(3-4):290--303, 2005.

\bibitem{Ismail2000}
M.~S. Ismail.
\newblock {A finite difference method for Korteweg-de Vries like equation with
  nonlinear dispersion}.
\newblock {\em International Journal of Computer Mathematics}, 74(2):185--193,
  2000.

\bibitem{Kim2007a}
P.~Kim.
\newblock {Invariantization of numerical schemes using moving frames}.
\newblock {\em BIT Numerical Mathematics}, 47:525--546, 2007.

\bibitem{Kim2008}
P.~Kim.
\newblock {Invariantization of the Crank-Nicolson method for Burgers'
  equation}.
\newblock {\em Physica D}, 237:243--254, 2008.

\bibitem{Korkmaz2010}
A.~Korkmaz.
\newblock {Numerical algorithms for solutions of Korteweg-de Vries equation}.
\newblock {\em Numerical Methods for Partial Differential Equations},
  26(6):1504--1521, 2010.

\bibitem{Lamb1980}
G.~L. Lamb.
\newblock {\em {Elements of soliton theory}}, volume~5.
\newblock Wiley, New York, 1980.

\bibitem{Leimkuhler2004}
B.~Leimkuhler and S.~Reich.
\newblock {\em {Simulating Hamiltonian Dynamics}}, volume~14 of {\em Cambridge
  Monographs on Applied and Computational Mathematics}.
\newblock Cambridge University Press, Cambridge, 2004.

\bibitem{Levy2004}
D.~Levy, C.-W. Shu, and J.~Yan.
\newblock {Local discontinuous Galerkin methods for nonlinear dispersive
  equations}.
\newblock {\em J. Comput. Phys.}, 196(2):751--772, 2004.

\bibitem{Lew2003}
A.~Lew, J.~Marsden, M.~Ortiz, and M.~West.
\newblock {An overview of variational integrators}.
\newblock In {\em Finite Element Methods: 1970s and beyond (CIMNE, 2003)},
  page~18, Barcelona, Spain, 2004.

\bibitem{Lin2006}
G.~Lin, L.~Grinberg, and G.~E. Karniadakis.
\newblock {Numerical studies of the stochastic Korteweg-de Vries equation}.
\newblock {\em Journal of Computational Physics}, 213(2):676--703, 2006.

\bibitem{Longuet-Higgins1963}
M.~S. Longuet-Higgins.
\newblock {The effect of non-linearities on statistical distributions in the
  theory of sea waves}.
\newblock {\em J. Fluid Mech.}, 17(03):459--480, March 1963.

\bibitem{Maday1988}
Y.~Maday and A.~Quarteroni.
\newblock {Error analysis for spectral approximation of the Korteweg-De Vries
  equation}.
\newblock {\em Mathematical Modelling And Numerical Analysis}, 22(3):499--529,
  1988.

\bibitem{Majda1997}
A.~J. Majda, D.~W. McLaughlin, and E.~G. Tabak.
\newblock {A one-dimensional model for dispersive wave turbulence}.
\newblock {\em Journal of Nonlinear Science}, 7(1):9--44, 1997.

\bibitem{Marsden1998}
J.~E. Marsden, G.~W. Patrick, and S.~Shkoller.
\newblock {Multisymplectic geometry, variational integrators, and nonlinear
  PDEs}.
\newblock {\em Communications in Mathematical Physics}, 199(2):52, 1998.

\bibitem{McLachlan1993}
R.~McLachlan.
\newblock {Symplectic integration of Hamiltonian wave equations}.
\newblock {\em Numerische Mathematik}, 66(1):465--492, 1993.

\bibitem{Miura1976}
R.~M. Miura.
\newblock {The Korteweg-de Vries equation: a survey of results}.
\newblock {\em SIAM Rev}, 18:412--459, 1976.

\bibitem{Miura1968}
R.~M. Miura, C.~S. Gardner, and M.~D. Kruskal.
\newblock {Korteweg-de Vries Equation and Generalizations. II. Existence of
  Conservation Laws and Constants of Motion}.
\newblock {\em Journal of Mathematical Physics}, 9(8):1204, 1968.

\bibitem{Moore2003}
B.~Moore and S.~Reich.
\newblock {Backward error analysis for multi-symplectic integration methods}.
\newblock {\em Numerische Mathematik}, 95(4):625--652, 2003.

\bibitem{Moore2003a}
B.~Moore and S.~Reich.
\newblock {Multi-symplectic integration methods for Hamiltonian PDEs}.
\newblock {\em Future Generation Computer Systems}, 19(3):395--402, 2003.

\bibitem{Olver1993}
P.~J. Olver.
\newblock {\em {Applications of Lie groups to differential equations}}, volume
  107 (2nd e of {\em Graduate Texts in Mathematics}.
\newblock Springer-Verlag, 1993.

\bibitem{Osborne1993}
A.~Osborne.
\newblock {Behavior of solitons in random-function solutions of the periodic
  Korteweg-de Vries equation}.
\newblock {\em Phys. Rev. Lett}, 71(19):3115--3118, November 1993.

\bibitem{Osborne2010}
A.~Osborne.
\newblock {\em {Nonlinear ocean waves and the inverse scattering transform}},
  volume~97.
\newblock Elsevier, 2010.

\bibitem{Osborne1995}
A.~R. Osborne.
\newblock {The numerical inverse scattering transform: nonlinear Fourier
  analysis and nonlinear filtering of oceanic surface waves}.
\newblock {\em Chaos Solitons Fractals}, 5(12):2623--2637, 1995.

\bibitem{Pelinovsky2006}
E.~Pelinovsky and A.~{Sergeeva (Kokorina)}.
\newblock {Numerical modeling of the KdV random wave field}.
\newblock {\em Eur. J. Mech. B/Fluids}, 25(4):425--434, July 2006.

\bibitem{Salupere2003}
A.~Salupere, J.~Engelbrecht, and P.~Peterson.
\newblock {On the long-time behaviour of soliton ensembles}.
\newblock {\em Math. Comp. Simul.}, 62(1-2):137--147, 2003.

\bibitem{Sanz-Sera1997}
J.-M. Sanz-Serna.
\newblock {Geometric integration}.
\newblock In I.~S. Duff and G.~A. Watson, editors, {\em The State of the Art in
  Numerical Analysis}, pages 121--143. Clarendon Press, Oxford, 1997.

\bibitem{Schiesser1994}
W.~E. Schiesser.
\newblock {Method of lines solution of the Korteweg-de vries equation}.
\newblock {\em Computers Mathematics with Applications}, 28(10-12):147--154,
  1994.

\bibitem{Schober2008}
C.~M. Schober and T.~H. Wlodarczyk.
\newblock {Dispersive properties of multisymplectic integrators}.
\newblock {\em J. Comput. Phys}, 227(10):5090--5104, May 2008.

\bibitem{Sergeeva2011}
A.~Sergeeva, E.~Pelinovsky, and T.~Talipova.
\newblock {Nonlinear random wave field in shallow water: variable Korteweg-de
  Vries framework}.
\newblock {\em Nat. Hazards Earth Syst. Sci.}, 11(2):323--330, February 2011.

\bibitem{Soderlind2003}
G.~S\"{o}derlind.
\newblock {Digital filters in adaptive time-stepping}.
\newblock {\em ACM Trans. Math. Software}, 29:1--26, 2003.

\bibitem{Soderlind2006}
G.~S\"{o}derlind and L.~Wang.
\newblock {Adaptive time-stepping and computational stability}.
\newblock {\em Journal of Computational and Applied Mathematics},
  185(2):225--243, 2006.

\bibitem{Taha1984}
T.~R. Taha and M.~J. Ablowitz.
\newblock {Analytical and numerical aspects of certain nonlinear evolution
  equations. III. Numerical, Korteweg-de Vries equation}.
\newblock {\em J. Comput. Phys}, 55(2):231--253, 1984.

\bibitem{Tayfun2007}
M.~A. Tayfun and F.~Fedele.
\newblock {Wave-height distributions and nonlinear effects}.
\newblock {\em Ocean Engineering}, 34(11-12):1631--1649, August 2007.

\bibitem{Trefethen2000}
L.~N. Trefethen.
\newblock {\em {Spectral methods in MatLab}}.
\newblock Society for Industrial and Applied Mathematics, Philadelphia, PA,
  USA, 2000.

\bibitem{Verner1978}
J.~H. Verner.
\newblock {Explicit Runge-Kutta methods with estimates of the local truncation
  error}.
\newblock {\em SIAM J. Num. Anal.}, 15(4):772--790, 1978.

\bibitem{Wang2003}
Y.~Wang, B.~Wang, and M.~Qin.
\newblock {Numerical Implementation of the Multisymplectic Preissman Scheme and
  Its Equivalent Schemes}.
\newblock {\em Applied Mathematics and Computation}, 149(2):299--326, 2003.

\bibitem{Zabusky1965}
N.~J. Zabusky and M.~D. Kruskal.
\newblock {Interaction of solitons in a collisionless plasma and the recurrence
  of initial states}.
\newblock {\em Phys. Rev. Lett}, 15:240--243, 1965.

\bibitem{Zahibo2009}
N.~Zahibo, E.~Pelinovsky, and A.~Sergeeva.
\newblock {Weakly damped KdV soliton dynamics with the random force}.
\newblock {\em Chaos, Solitons \& Fractals}, 39(4):1645--1650, February 2009.

\bibitem{Zakharov1971}
V.~E. Zakharov.
\newblock {Kinetic Equation for Solitons}.
\newblock {\em Sov. Phys. - JETP}, 60:993--1000, 1971.

\bibitem{Zakharov1972a}
V.~E. Zakharov and L.~D. Faddeev.
\newblock {Korteweg-de Vries equation: A completely integrable Hamiltonian
  system}.
\newblock {\em Functional Analysis and Its Applications}, 5(4):280--287, 1972.

\bibitem{Zakharov2001}
V.~E. Zakharov, P.~Guyenne, A.~N. Pushkarev, and F.~Dias.
\newblock {Wave turbulence in one-dimensional models}.
\newblock {\em Physica D}, 153:573--619, 2001.

\bibitem{Zakharov1992}
V.~E. Zakharov, V.~S. Lvov, and G.~Falkovich.
\newblock {\em {Kolmogorov Spectra of Turbulence I}}.
\newblock Springer-Verlag, Berlin, 1992.

\bibitem{Zhao2000}
P.~F. Zhao and M.~Z. Qin.
\newblock {Multisymplectic geometry and multisymplectic Preissmann scheme for
  the KdV equation}.
\newblock {\em Journal of Physics A: Mathematical and General},
  33(18):3613--3626, May 2000.

\end{thebibliography}
\bibliographystyle{plain}

\end{document}